\documentclass[twocolumn]{aastex61}
\pdfoutput=1
\usepackage{amsmath}
\usepackage{url}

\newcommand{\msun} {M_\odot}
\newcommand{\lsun} {L_\odot}
\newcommand{\Teff} {T_{\rm eff}}
\newcommand\gta{\lower 0.5ex\hbox{$\ \buildrel > \over \sim\ $}}
\newcommand\lta{\lower 0.5ex\hbox{$\ \buildrel < \over \sim\ $}}

\begin{document}

\title{On the Spectral Evolution of Hot White Dwarf Stars. \ II. Time-dependent \\ Simulations of Element Transport in Evolving White Dwarfs with STELUM \vspace*{2mm}}

\author{A. B\'edard}
\affiliation{D\'epartement de Physique, Universit\'e de Montr\'eal, Montr\'eal, QC H3C 3J7, Canada; email: antoine.bedard@umontreal.ca}

\author{P. Brassard}
\affiliation{D\'epartement de Physique, Universit\'e de Montr\'eal, Montr\'eal, QC H3C 3J7, Canada; email: antoine.bedard@umontreal.ca}

\author{P. Bergeron}
\affiliation{D\'epartement de Physique, Universit\'e de Montr\'eal, Montr\'eal, QC H3C 3J7, Canada; email: antoine.bedard@umontreal.ca}

\author{S. Blouin}
\affiliation{Los Alamos National Laboratory, Los Alamos, NM 87545, USA}
\affiliation{Department of Physics and Astronomy, University of Victoria, Victoria, BC V8W 2Y2, Canada \vspace*{8mm}}

\shorttitle{Simulations of Transport in White Dwarfs}
\shortauthors{B\'edard, Brassard, Bergeron \& Blouin}

\begin{abstract}

White dwarf stars are subject to various element transport mechanisms that can cause their surface composition to change radically as they cool, a phenomenon known as spectral evolution. In this paper, we undertake a comprehensive theoretical investigation of the spectral evolution of white dwarfs. First, we introduce STELUM, a new implementation of the stellar evolutionary code developed at the Universit\'e de Montr\'eal. We provide a thorough description of the physical content and numerical techniques of the code, covering the treatment of both stellar evolution and chemical transport. Then, we present two state-of-the-art numerical simulations of element transport in evolving white dwarfs. Atomic diffusion, convective mixing, and mass loss are considered simultaneously as time-dependent diffusive processes and are fully coupled to the cooling. We first model the PG 1159$-$DO$-$DB$-$DQ evolutionary channel: a helium-, carbon-, and oxygen-rich PG 1159 star transforms into a pure-helium DB white dwarf due to gravitational settling, and then into a helium-dominated, carbon-polluted DQ white dwarf through convective dredge-up. We also compute for the first time the full DO$-$DA$-$DC evolutionary channel: a helium-rich DO white dwarf harboring residual hydrogen becomes a pure-hydrogen DA star through the float-up process, and then a helium-dominated, hydrogen-bearing DC star due to convective mixing. We demonstrate that our results are in excellent agreement with available empirical constraints. In particular, our DO$-$DA$-$DC simulation perfectly reproduces the lower branch of the bifurcation observed in the Gaia color-magnitude diagram, which can therefore be interpreted as a signature of spectral evolution. \vspace*{2mm}

\end{abstract}

\section{Introduction} \label{sec:intro}

In the course of their evolution, stars are subject to several physical mechanisms that may alter their chemical composition. Besides the obvious example of nuclear burning, a number of processes give rise to the transport of matter within a star and thereby change the distribution of the elements with depth. These transport mechanisms belong to one of two categories: microscopic processes, which influence individual chemical constituents differentially, and macroscopic processes, which affect the stellar material as a whole (see \citealt{salaris2017} for a recent review).

Microscopic transport, often referred to as atomic diffusion, is the net result of interparticle collisions in the presence of a gradient of some physical quantity. Three types of diffusion can be distinguished: chemical diffusion, gravitational settling, and thermal diffusion, which are induced by composition, pressure, and temperature gradients, respectively. On one hand, both gravitational settling and thermal diffusion tend to separate the atomic species according to their mass, with heavier ions moving toward regions of higher pressure and temperature (that is, toward the center of a star). On the other hand, chemical diffusion tends to homogenize the composition and thus to counteract the above-mentioned effect \citep{thoul1994,michaud2015}. Another form of selective transport is radiative levitation, whereby photons transfer part of their net outward momentum to ions of a given type. This phenomenon also acts against gravitational sedimentation, especially in the outer layers of a hot star \citep{chayer1995,michaud2015}.

As for macroscopic transport, convection is the most common mechanism driving large-scale flows of matter. Besides carrying energy very efficiently, these rapid motions lead to the complete mixing of the various chemical species, thereby impeding the effects of atomic diffusion \citep{kippenhahn2012}. In addition to ordinary convection, phenomena such as convective overshoot, semi-convection, and thermohaline convection provide additional sources of mixing under certain circumstances \citep{kippenhahn1980,langer1985,freytag1996}. Stellar winds and the accompanying mass loss constitute another example of macroscopic transport. In particular, at the surface of a hot star, the radiation field can be strong enough to power a continuous, undifferentiated outflow of material. The impact of such a wind is to slow down atomic diffusion in the outer envelope \citep{kudritzki2000,unglaub2000}. Obviously, the reverse mechanism, mass accretion, either from the interstellar medium or from a companion, can also influence the surface composition \citep{alcock1980,koester2009}.

White dwarf stars provide a unique window on chemical transport in an astrophysical context. In these compact, burnt-out, slowly cooling stellar remnants, gravitational settling is expected to be dominant \citep{schatzman1958,paquette1986b} and thus to produce a stratified chemical structure: a carbon/oxygen core, surrounded by a pure-helium mantle, itself surrounded by a pure-hydrogen layer. While pure-hydrogen atmosphere white dwarfs (type DA\footnote{The coolest hydrogen-atmosphere white dwarfs actually belong to the DC spectral class since hydrogen lines disappear below $\Teff \sim 5000$ K.}) are indeed ubiquitous, atmospheres made of helium (types DO, DB, DC) and/or polluted by trace elements (types DBA, DQ, DZ, and many others) are also observed in significant numbers (\citealt{kepler2019} and references therein). Furthermore, a wealth of evidence suggests that the surface composition of white dwarfs can change as they cool, a phenomenon known as spectral evolution (\citealt{bedard2020}, hereafter Paper I, and references therein). Therefore, it is clear that gravitational settling is not the sole transport process at work in degenerate stars.

First and foremost, the very existence of hydrogen-deficient white dwarfs is particularly telling. Stellar remnants are expected to retain a ``thick'' superficial hydrogen layer of fractional mass $q_{\rm H} \equiv M_{\rm H}/M \sim 10^{-4}$ \citep{iben1984,renedo2010}. The fact that one in four objects enters the cooling sequence with a hydrogen-deficient surface (Paper I) then implies that radical chemical alterations take place in pre-white dwarf evolutionary phases. It is believed that most of these DO white dwarfs descend from helium-, carbon-, and oxygen-rich PG 1159 stars, which themselves are the result of a late helium-shell flash \citep{althaus2005a,werner2006}. The hydrogen deficiency is a direct consequence of this event: the flash produces an extensive convection zone, which causes the superficial hydrogen to be mixed deeply into the stellar envelope and hence almost entirely burned.

At the beginning of the cooling sequence, the idea that gravitational settling leaves only pure-hydrogen and pure-helium atmospheres does not stand up to scrutiny. For instance, some PG 1159 stars keep their surface carbon and oxygen down to an effective temperature $\Teff \sim 75,000$ K \citep{werner2006}. In addition, several hot hydrogen-rich DA white dwarfs exhibit atmospheric traces of helium and/or metals (\citealt{gianninas2010,barstow2014}; Paper I). This indicates that the downward diffusion of heavy elements is initially hampered by a competing transport mechanism. The most likely candidate is a weak radiative stellar wind \citep{unglaub1998,unglaub2000,quirion2012}, although radiative levitation is also thought to play some role \citep{chayer1995,dreizler1999}.

Below $\Teff \sim 75,000$ K, as stellar winds fade and gravitational settling starts to operate, another remarkable feature emerges: the fraction of helium-dominated white dwarfs gradually declines with decreasing effective temperature, implying that these objects can become hydrogen-rich as they cool. More specifically, two-thirds of all hot DO stars turn into DA stars before they reach $\Teff \sim 30,000$ K (Paper I). This transformation is usually interpreted as the result of the so-called float-up process: a small amount of residual hydrogen, initially diluted into the helium envelope, is slowly carried upward by diffusion, ultimately forming a hydrogen shell at the surface \citep{fontaine1987,althaus2005b}. This scenario is further supported by the detection, around $\Teff \sim 40,000-50,000$ K, of white dwarfs with extremely thin superficial hydrogen layers ($q_{\rm H} \sim 10^{-17}-10^{-16}$), which represent an intermediate stage in the DO-to-DA transition (\citealt{manseau2016}; Paper I).

Going further down the effective temperature scale, element transport produces another striking observational signature: the fraction of helium-dominated white dwarfs re-increases below $\Teff \sim 20,000$ K \citep{ourique2019,blouin2019a,genest-beaulieu2019,cunningham2020}. This trend suggests that DA stars with newly formed thin hydrogen layers transform back into helium-atmosphere objects. The mechanism at the source of this DA-to-DB/DC spectral metamorphosis is convection. In fact, two similar but distinct phenomena, referred to in the literature as convective dilution and convective mixing, are believed to occur, respectively above and below $\Teff \sim 15,000$ K.

In the first case, the amount of hydrogen accumulated at the surface is small enough ($q_{\rm H} \lta 10^{-14}$) to allow the development of a convection zone in the underlying helium mantle. As the effective temperature decreases, the convective region grows and erodes the superficial hydrogen layer from below, until the latter is completely diluted into the helium envelope. The observational outcome is a helium-dominated atmosphere with a minute trace of hydrogen, namely, a DBA white dwarf \citep{macdonald1991,rolland2018,rolland2020}.

In the second case, the superficial hydrogen shell is thicker ($q_{\rm H} \gta 10^{-14}$) and the helium envelope remains convectively stable. Instead, it is in the hydrogen layer that a convection zone appears. When the advective motions reach into the underlying helium, the latter is massively dredged-up to the surface, such that the hydrogen ends up being thoroughly mixed in a helium-dominated outer envelope. Although still detectable immediately after the mixing episode, the hydrogen quickly becomes invisible as the effective temperature decreases. Consequently, the white dwarf evolves briefly into a helium-rich DA star, and then into a DC star \citep{macdonald1991,chen2011,rolland2018}. An analogous phenomenon is thought to be responsible for the existence of carbon-polluted DQ white dwarfs below $\Teff \sim 10,000$ K: these objects descend from DB stars in which settling carbon (a relic of their PG 1159 past) is dredged-up from the deep envelope to the surface by the growing helium convection zone \citep{pelletier1986,dufour2005,camisassa2017}.

Finally, 25$-$50\% of cool white dwarfs show atmospheric contamination by heavy elements, especially calcium, magnesium, and iron \citep{koester2014,coutu2019}. It is now firmly established that these DAZ, DBZ, and DZ stars owe their spectral appearance to the accretion of tidally disrupted asteroids or planets \citep{farihi2010,jura2014}. Since such rocky material is expected to sink out of sight on extremely short timescales \citep{dupuis1992,koester2009}, the existence of so many metal-polluted white dwarfs reveals that this type of accretion is quite frequent.

The above considerations have made evident that the spectral evolution of white dwarfs is governed by a complex interplay between numerous transport mechanisms. This topic has been extensively studied from an empirical perspective in recent years, largely thanks to the Sloan Digital Sky Survey \citep{york2000} and the Gaia mission \citep{gaia2016}. Nevertheless, our theoretical understanding of the dynamical phenomena giving rise to spectral evolution remains scarce. Several works have attempted to model chemically evolving white dwarfs, but the vast majority of them adopted a crude treatment of chemical transport, either assuming diffusive equilibrium between the various processes or ignoring the feedback of composition changes on the evolution. Detailed evolutionary calculations in which time-dependent transport is fully coupled to white dwarf cooling have seldom been carried out. Moreover, conflicting results have sometimes been reported by different groups from such self-consistent calculations. In this context, we feel that improvements in the modeling of the spectral evolution are greatly needed. This endeavor not only constitutes a remarkable opportunity to further our understanding of element transport in astrophysical objects, but is also of crucial relevance for the fields of cosmochronology and asteroseismology, given that the chemical history of degenerate stars impact their cooling and pulsation properties \citep{fontaine2001,fontaine2008}.

The present paper is the second of a series dedicated to the spectral evolution of white dwarfs. In Paper I, we performed a comprehensive spectroscopic analysis of a large sample of hot white dwarfs in order to improve our empirical knowledge of this phenomenon. In the present work, we turn to the theoretical side of the subject and undertake an extended modeling study of spectral evolution. First, we provide an exhaustive description of our stellar evolutionary code in Section \ref{sec:code}. Then, we present and discuss the results of our time-dependent simulations of chemical transport in evolving white dwarf models in Section \ref{sec:simu}. Finally, our conclusions are summarized in Section \ref{sec:conclu}.

\section{The STELUM Evolutionary Code} \label{sec:code}

The Montreal white dwarf evolutionary code was introduced some 20 years ago by \citet{fontaine2001} and subsequently updated by \citet{dufour2005}, \citet{quirion2012}, \citet{van-grootel2013}, and \citet{brassard2015}, among others. In recent years, one of us (P. Brassard) has devoted considerable efforts to the development of a new, state-of-the-art version of the code, completely rewritten in modern Fortran and in modular form, with substantially improved capabilities. This revamped implementation of our evolutionary code is named STELUM, for STELlar modeling from the Universit\'e de Montr\'eal. STELUM is specifically but not exclusively designed for the modeling of white dwarfs and hot subdwarfs. It can build and evolve complete stellar models, down from the center up to the very surface, in which the transport of chemical species is fully coupled to the evolution. STELUM has already been used in a few works (\citealt{rolland2020,blouin2020b,blouin2020a}; Paper I), but has never been properly described. The goal of the present section is to remedy this situation. In the following, we provide a detailed description of the code, with a particular focus on the treatment of element transport.

\subsection{Stellar Structure and Evolution} \label{sec:code_evol}

STELUM relies on the standard one-dimensional theory of stellar structure and evolution (see Appendix \ref{sec:app_a_evol}). Before addressing the physical content of the code, a few aspects of our numerical approach are worth discussing. The fundamental equations are not solved directly in their conventional Lagrangian form; rather, they are first recast in terms of a new independent depth variable. This strategy is motivated by our wish to compute high-precision stellar models down from the center up to the very surface (which is essential to simulate the spectral evolution of white dwarfs properly). While the usual interior mass $m$ is an appropriate choice of independent variable across most of the stellar interior, the quantity $\log q$, where $q = 1 - m/M$ is the exterior mass, is better suited to describe the outermost layers. In order to model both regimes accurately, STELUM adopts the independent mass variable $\xi$ defined by Equations 1$-$3 of \citet{tassoul1990}, which we reproduce here for completeness:
\begin{equation}
q =
\begin{cases}
1 + c_1 \xi^3 , & 0.0 \le \xi \le 0.4 , \\
c_2 + c_3 \xi + c_4 \xi^2 + c_5 \xi^3 , & 0.4 \le \xi \le 1.0 , \\
c_6 \exp{ \left[ 15 \left( 1 - \xi \right) \right] } , & \xi \ge 1.0 ,
\end{cases}
\label{eq:xi_to_q}
\end{equation}
where the values of the coefficients are $c_1 = -2.1825397$, $c_2 = 0.64021163$, $c_3 = 2.6984127$, $c_4 = -6.7460317$, $c_5 = 3.4391534$, and $c_6 = 0.03174603$. With this definition, $\xi \propto m^{1/3}$ in the deep interior and $\xi \propto - \log q$ in the outermost layers, as needed.

The numerical solution of the modified structure equations is then carried out using the so-called Galerkin finite-element method, with either linear, quadratic, or cubic elements. The technique is described at length in \citet{brassard1992} and the details will not be repeated here. Nevertheless, we want to stress that the use of a finite-element scheme to build and evolve stellar models is perhaps the most important and defining feature of STELUM. Indeed, as demonstrated by \citet{brassard1992} in the context of stellar pulsations, this method outperforms the various finite-difference schemes employed by most evolution and pulsation codes in terms of stability, accuracy, and effectiveness. Furthermore, STELUM relies on an adaptive time-step algorithm, whereby the time step between two successive models of an evolutionary sequence is selected according to a pre-specified tolerance for the relative changes in radius, pressure, and temperature averaged over the star.

A number of functions appearing in the basic structure equations must be specified a priori through an appropriate description of the microphysics of the stellar material. This so-called constitutive physics comprises the equation of state, the Rosseland mean opacity, the temperature gradient, and the rates of energy gain/loss due to nuclear reactions, neutrino emission, and gravothermal processes (see Appendix \ref{sec:app_a_evol}). In the rest of this section, we give an exhaustive account of how these physical ingredients are handled in STELUM.

The equation of state consists of four tables of thermodynamic data suitable for pure hydrogen, helium, carbon, and oxygen compositions. In order to cover a large domain of the density-temperature plane, each table is made of three parts corresponding to three different physical regimes, connected as smoothly as possible. In the low-density regime, the thermodynamic properties of the partially ionized, non-degenerate, almost ideal gas are obtained by solving the appropriate network of Saha equations including a Coulomb correction term. In intermediate-density regions where partial ionization, partial electron degeneracy, and non-ideal effects all occur, we use the equation of state of \citet{saumon1995} for hydrogen and helium, an improved version of the equation of state of \citet{fontaine1977} for carbon, and an unpublished equation of state for oxygen (developed by the late G. Fontaine at the Universit\'e de Montr\'eal). Finally, in the high-density regime, the state of the completely ionized, strongly electron-degenerate, highly non-ideal plasma is described according to the formalism introduced by \citet{lamb1974} and improved by \citet{kitsikis2005}, which accounts for both liquid and solid (bcc) phases. For elements heavier than oxygen, we employ the oxygen equation of state scaled to the appropriate atomic weight and charge. For mixtures of multiple species, the thermodynamic data are interpolated in composition following the additive-volume prescription of \citet{fontaine1977}.

The total opacity is a combination of two contributions, respectively associated with radiative and conductive heat transfer processes. The default radiative opacities are the OPAL data for pure substances \citep{iglesias1996}, supplemented at low temperatures with our own hydrogen and helium opacities computed using the model-atmosphere code described in \citet{bergeron1995,bergeron1997,bergeron2001}. For mixtures of multiple species, the pure-element opacities are linearly interpolated in composition; note that the OPAL tables for arbitrary mixtures are also included in STELUM but are not used by default, because they are very time-consuming and of little consequence in most applications. Alternatively, the analogous OP data \citep{badnell2005} and the low-temperature opacities of \citet{ferguson2005} are also available. As for the conductive opacities, we rely on the data of \citet{cassisi2007} and the corrections of \citet{blouin2020a} in the moderately coupled and moderately degenerate regime. Other available options are the older opacities of \citet{hubbard1969} and \citet{itoh1983,itoh1984}.

The temperature gradient takes different forms depending on the presence or absence of convection, as determined by the standard Schwarzschild criterion. On one hand, a convectively stable medium is characterized by the radiative temperature gradient,
\begin{equation}
\nabla_{\rm rad} = \frac{3 W l P \kappa}{64 \pi \sigma G m T^4} ,
\label{eq:rad_grad}
\end{equation}
where $\sigma$ is the Stefan-Boltzmann constant, $G$ is the gravitational constant, $l$ is the luminosity, $P$ is the total pressure, $T$ is the temperature, $\kappa$ is the opacity, and $W$ is a correction factor accounting for the non-diffusive heat transfer at small optical depths (see below). On the other hand, a convectively unstable region is characterized by the convective temperature gradient, $\nabla_{\rm conv}$, which must be obtained from a model of convection. The default option is the so-called ML2 version of the mixing-length theory, in which the mixing-length parameter is $\alpha=1.0$ \citep{bohm-vitense1958,tassoul1990}. Similarly to the above expression for $\nabla_{\rm rad}$, we incorporate the $W$ factor into the usual equations for $\nabla_{\rm conv}$ in order to consider non-diffusive effects near the surface (see Appendix \ref{sec:app_a_conv}). Furthermore, note that the alternative theory of stellar convection introduced by \citet{canuto1991} is also implemented in STELUM.

The atmospheric correction parameter $W$ is inspired from the formalism of \citet{henyey1965}. At the present time, it is evaluated under the assumption of a simple grey atmosphere, in which case it takes the following analytic form:
\begin{equation}
W = 1 + \frac{dH}{d \tau} ,
\label{eq:atmo_fac}
\end{equation}
where $\tau$ is the Rosseland optical depth and $H$ is the Hopf function. The latter is approximated with a seventh-order polynomial fit to a numerical solution of its defining integral equation (see Appendix \ref{sec:app_a_hopf}). We are currently working on implementing in STELUM a more realistic approach, which consists in extracting $W$ from the detailed non-grey model atmospheres computed by our group.

The nuclear energy generation rate is defined as the sum of the energy produced per unit mass and unit time by all relevant nuclear reactions. To calculate this, one must supply a reaction network along with individual reaction cross sections and electron screening factors (see Appendix \ref{sec:app_a_nuc}). The nuclear network of STELUM incorporates the main reactions involved in hydrogen burning (both the PP chain and the CNO cycle) and helium burning. The burning of heavier elements is not presently included. There are three possible levels of sophistication regarding the species tracked in the calculations: \\
$\bullet$ a basic network: H, He, C, O; \\
$\bullet$ a standard network: $^{1}$H, $^{2}$H, $^{3}$He, $^{4}$He, $^{12}$C, $^{13}$C, $^{14}$N, $^{15}$N, $^{16}$O, $^{17}$O, $^{20}$Ne; \\
$\bullet$ an extended network: $^{1}$H, $^{2}$H, $^{3}$He, $^{4}$He, $^{7}$Be, $^{7}$Li, $^{8}$B, $^{12}$C, $^{13}$C, $^{13}$N, $^{14}$N, $^{15}$N, $^{15}$O, $^{16}$O, $^{17}$O, $^{18}$O, $^{17}$F, $^{18}$F, $^{19}$F, $^{20}$Ne. \\
By default, the reaction cross sections are taken from \citet{angulo1999}, but the data of \citet{caughlan1988} and \citet{cyburt2010} are also available. The electron screening factors are computed following the prescription of \citet{dewitt1973} and \citet{graboske1973}.

Similar considerations apply to the neutrino energy loss rate (see Appendix \ref{sec:app_a_nuc}). In this case, the leptonic reactions included in STELUM are the pair, photo-, plasma, bremsstrahlung, and recombination neutrino processes. For all these mechanisms, the rates of energy loss are taken from \citet{itoh1996}.

Finally, the gravothermal energy generation rate is defined as the rate at which heat is removed from the stellar material due to changes in its thermodynamic properties. This quantity is a function of time $t$ and is formulated in STELUM as
\begin{equation}
\begin{split}
& \epsilon_{\rm grav} = \frac{P}{\rho} \frac{\chi_T}{\chi_\rho} \left( \frac{d \ln P}{dt} - \frac{1}{\nabla_{\rm ad}} \frac{d \ln T}{dt} \right) \\
& - \sum_{i=1}^{I} \left( u_i + \frac{P}{\rho_i} \right) \frac{dX_i}{dt} + l_{\rm cr} \frac{dm_{\rm cr}}{dt} \ \delta \left( m - m_{\rm cr} \right) .
\end{split}
\end{equation}
In this expression, $\rho$ is the total mass density, $\chi_T$ and $\chi_\rho$ are standard thermodynamic derivatives, $\nabla_{\rm ad}$ is the adiabatic temperature gradient, $X_i$, $\rho_i$, and $u_i$ respectively denote the mass fraction abundance, mass density, and internal energy per unit mass of element $i$, and $I$ stands for the number of chemical species considered. Besides, note that all time derivatives here and in the rest of the paper are Lagrangian derivatives. The terms involving the derivatives of the pressure and temperature represent the well-known contribution of the expansion/contraction of the star to the energy budget. The terms involving the derivatives of the elemental mass fractions account for changes in internal energy due to changes in chemical composition. Here, $dX_i/dt$ must be deduced from the physics of particle transport, which is discussed extensively in Section \ref{sec:code_trans}.

The last term applies only to crystallizing white dwarfs and represents the energy generated by core crystallization: $l_{\rm cr}$ is the energy released per unit mass, $m_{\rm cr}$ is the mass coordinate of the solidification front, and the $\delta$ function indicates that the energy is deposited at $m = m_{\rm cr}$. The crystallized mass $m_{\rm cr}$ is determined by the usual condition on the Coulomb coupling parameter,
\begin{equation}
\Gamma = \frac{\overline{Z^{5/3}} e^2}{a_e k_{\rm B} T} \ge \Gamma_{\rm cr} ,
\end{equation}
where $e$ is the elementary charge, $k_{\rm B}$ is the Boltzmann constant, $\overline{Z^{5/3}}$ is the average of $Z_i^{5/3}$ over all species (with $Z_i$ being the charge of ion $i$), $a_e = ( 3 / 4 \pi n_e )^{1/3}$ is the electron-sphere radius (with $n_e$ denoting the electron number density), and $\Gamma_{\rm cr}$ is the value of $\Gamma$ above which crystallization occurs. The latter quantity, which depends on the local chemical composition, is obtained from the carbon/oxygen phase diagram of \citet{blouin2021}. In principle, $l_{\rm cr}$ comprises only the latent heat released by the phase transition and is simply given by
\begin{equation}
l_{\rm cr} = 0.77 \frac{N_{\rm A} k_{\rm B} T}{\mu_{\rm ion}} ,
\end{equation}
where $N_{\rm A}$ is Avogadro's number, $\mu_{\rm ion}$ is the mean molecular weight per ion, and the numerical factor is taken from \citet{salaris2000}. However, in practice, we follow \citet{isern2000} and add a second term designed to account for the gravitational energy produced by the phase separation of carbon and oxygen upon crystallization. Using Equations 2 and 8 of \citet{isern2000} and assuming a binary mixture of carbon and oxygen, the previous expression becomes
\begin{equation}
\begin{split}
l_{\rm cr} = \left[ 0.77 + 0.31 \mu_{\rm ion} \left( \Gamma_e - \frac{ \langle T \Gamma_e \rangle }{T} \right) \Delta X_{\rm O} \right] \frac{N_{\rm A} k_{\rm B} T}{\mu_{\rm ion}} ,
\end{split}
\end{equation}
where $\Gamma_e = \Gamma / \overline{Z^{5/3}}$ is the electron coupling parameter, $\Delta X_{\rm O}$ is the difference in oxygen mass fraction between the solid and liquid phases, and the average $\langle T \Gamma_e \rangle$ is taken over the Rayleigh-Taylor-unstable liquid region just above the crystallization front (see \citealt{isern1997}). We again rely on the phase diagram of \citet{blouin2021} to evaluate the degree of oxygen enrichment in the solid core $\Delta X_{\rm O}$\footnote{The improvements of \citet{blouin2020a} to the conductive opacities and of \citet{blouin2021} to the treatment of phase separation were incorporated into STELUM only after the evolutionary sequences presented in Paper I and in this work were computed. Thus, these calculations rely on the uncorrected conductive opacities of \citet{cassisi2007} and ignore the energy released through phase separation (only the latent heat is included).}.

\subsection{Element Transport} \label{sec:code_trans}

STELUM incorporates a detailed treatment of element transport, as required to model the chemical evolution of white dwarfs. Several transport phenomena can be considered simultaneously: chemical diffusion, gravitational settling, thermal diffusion, stellar winds, external accretion, ordinary convection, convective overshoot, semi-convection, and thermohaline convection. These mechanisms are all treated as time-dependent diffusive processes, and the feedback of composition changes on the evolution of the star is fully taken into account. As the central purpose of the present paper is to exhibit and exploit these capabilities, we believe it is appropriate to elaborate on the relevant details here.

The time evolution of the elemental mass fraction $X_i$ at a given radius $r$ is described by the following transport equation:
\begin{equation}
\begin{split}
& \frac{dX_i}{dt} = S_{{\rm nuc},i} \\
& - \frac{1}{r^2 \rho} \frac{d}{dr} \left( r^2 \rho \left[ \left( v_i + v_{\rm wind} + v_{\rm acc} \right) X_i - D \frac{dX_i}{dr} \right] \right) ,
\end{split}
\label{eq:trans}
\end{equation}
where $S_{{\rm nuc},i}$ is a source/sink term accounting for the creation/destruction of element $i$ by nuclear reactions (see Appendix \ref{sec:app_a_nuc}), $v_i$ is the diffusion velocity of element $i$, $v_{\rm wind}$ is the stellar wind velocity, $v_{\rm acc}$ is the accretion velocity, and $D$ is the mixing coefficient (also called the macroscopic diffusion coefficient). As before, the time derivative is a Lagrangian derivative. The full chemical evolution is governed by a set of $I$ such equations, one for each atomic species. This form of the transport equation nicely emphasizes the distinction between microscopic and macroscopic transport: the effects of atomic diffusion are contained in $v_i$, whereas the effects of large-scale motions are captured by $v_{\rm wind}$, $v_{\rm acc}$, and $D$. In the following, we succinctly outline the physics behind each of these terms.

The diffusion velocities are obtained from the formalism of \citet{burgers1969} describing atomic diffusion in a multi-component fluid. For a mixture of $I+1$ species ($I$ elements plus the free electrons), one must solve a system of $2I+4$ coupled equations, comprising $I+1$ equations of momentum conservation,
\begin{equation}
\begin{split}
& \frac{dP_i}{dr} + \rho_i \left( g - g_{{\rm rad,}i} \right) - n_i Z_i e E = \\
& \sum_{j \ne i}^{I+1} K_{ij} \left( v_j - v_i \right) + \sum_{j \ne i}^{I+1} K_{ij} z_{ij} \frac{A_j r_i - A_i r_j}{A_i + A_j} ,
\end{split}
\label{eq:diff_1}
\end{equation}
$I+1$ equations of energy conservation,
\begin{equation}
\begin{split}
& \frac{5}{2} n_i k_{\rm B} T \frac{d \ln T}{dr} = \\
& - \frac{2}{5} K_{ii} z_{ii}'' r_i - \frac{5}{2} \sum_{j \ne i}^{I+1} \frac{K_{ij} z_{ij} A_j}{A_i + A_j} \left( v_j - v_i \right) \\
& - \sum_{j \ne i}^{I+1} \frac{K_{ij}}{\left( A_i + A_j \right)^2} \left( 3 A_i^2 + z_{ij}' A_j^2 + \frac{4}{5} z_{ij}'' A_i A_j \right) r_i \\
& + \sum_{j \ne i}^{I+1} \frac{K_{ij} A_i A_j}{\left( A_i + A_j \right)^2} \left( 3 + z_{ij}' - \frac{4}{5} z_{ij}'' \right) r_j ,
\end{split}
\label{eq:diff_2}
\end{equation}
and two additional equations imposing the constraints of no net mass flow and no net electric current (see for instance \citealt{salaris2017}). Note that the index $I+1$ refers to the electrons. In these expressions, $E$ is the electric field, $g$ is the gravitational acceleration, $A_i$, $P_i$, $n_i$, $g_{{\rm rad,}i}$, and $r_i$ are the atomic weight, partial pressure, number density, radiative acceleration, and residual heat flow of species $i$, respectively, and $K_{ij}$, $z_{ij}$, $z_{ij}'$, and $z_{ij}''$ are the so-called resistance coefficients, which capture the dynamics of interparticle collisions. All other symbols have been defined previously. It is instructive to discuss very briefly the physical meaning of the various terms in Equation \ref{eq:diff_1}. The left-hand side is the total force per unit volume on particles of type $i$ due to the partial pressure gradient, the gravitational field, and the electric field. These contributions collectively give rise to the processes referred to as chemical diffusion and gravitational settling. In the second term, the phenomenon of radiative levitation is accounted for by employing an effective (reduced) gravity $g - g_{{\rm rad,}i}$. Finally, the last term involving the residual heat flows on the right-hand side represents the contribution of thermal diffusion.

A number of quantities appearing in the Burgers formalism must be specified a priori to close the system. The partial pressure $P_i$ must be expressed in terms of the number density $n_i$, which is often done through an ideal gas law. Because this approximation breaks down in white dwarf interiors, we add the Coulomb correction term proposed by \citet{beznogov2013}, so that the partial pressure gradient becomes
\begin{equation}
\begin{split}
\frac{dP_i}{dr} = n_i k_{\rm B} T \left( \frac{d \ln n_i}{dr} + \frac{d \ln T}{dr} \right) - \frac{3}{10} \frac{Z_i^{5/3} e^2}{a_e} \frac{d\ln n_e}{dr} n_i ,
\end{split}
\label{eq:diff_3}
\end{equation}
where $n_e \equiv n_{I+1}$. A prescription for the calculation of the radiative acceleration $g_{{\rm rad,}i}$ is also needed. At the present time, this feature is not included in STELUM (we set $g_{{\rm rad,}i} = 0$ for all $i$). Finally, the resistance coefficients $K_{ij}$, $z_{ij}$, $z_{ij}'$, and $z_{ij}''$ are linked to the so-called collision integrals, which must also be known. These are evaluated from the results of \citet{fontaine2015}, which are based on an improved version of the method described at length in \citet{paquette1986a}. With these quantities specified, the Burgers equations contain $2I+3$ unknown variables: the $I+1$ diffusion velocities $v_i$, the $I+1$ residual heat flows $r_i$, and the electric field $E$. Given that the full system comprises $2I+4$ equations, the problem is over-constrained. A common approach is to consider the gravitational acceleration $g$ as an additional unknown variable and to compare the derived value with the expected value $g = G m / r^2$ as a consistency check \citep{thoul1994}. However, this strategy is problematic when diffusion takes place in electron-degenerate material, such as in white dwarfs. The issue is that Equation \ref{eq:diff_1}, although appropriate for the non-degenerate ions, cannot be applied to the degenerate electrons, because their partial pressure does not follow a simple analytic formula such as Equation \ref{eq:diff_3}. A straightforward solution is to drop Equation \ref{eq:diff_1} for the electrons and to treat $g = G m / r^2$ as a known parameter, leaving a closed set of $2I+3$ equations and $2I+3$ unknown variables \citep{pelletier1986,paxton2018}. This is the approach adopted in STELUM\footnote{As pointed out by \citet{paxton2018}, another issue in the application of the Burgers formalism to white dwarfs is that the expression on the left-hand side of Equation \ref{eq:diff_2} implicitly assumes an ideal, non-degenerate gas. However, this inconsistency is expected to be of little consequence because it only affects the thermal-diffusion terms, which are generally much smaller than the gravitational-settling terms in white dwarfs.}.

The wind velocity follows from simple mass conservation arguments. Assuming that the wind is weak enough so that the stellar mass can be considered constant, the wind velocity is given by
\begin{equation}
v_{\rm wind} = - \frac{\dot{M}_{\rm wind}}{4 \pi r^2 \rho} ,
\end{equation}
where the mass-loss rate $\dot{M}_{\rm wind}$ must be specified. The mass-loss rate is inherently negative, so that the wind velocity is positive (that is, the stellar material is pushed outward). Because mass accretion is the mathematical opposite of mass loss, an identical expression applies to the accretion velocity,
\begin{equation}
v_{\rm acc} = - \frac{\dot{M}_{\rm acc}}{4 \pi r^2 \rho} ,
\end{equation}
where the accretion rate $\dot{M}_{\rm acc}$ must also be provided. The accretion velocity is directed inward since $\dot{M}_{\rm acc}$ is by definition a positive quantity.

The macroscopic transport coefficient $D$ takes different forms depending on the locally dominant mixing mechanism. In a convection zone, we adopt the mixing coefficient prescribed by \citet{langer1985},
\begin{equation}
D_{\rm conv} = \frac{1}{3} v_{\rm conv} \ell_{\rm conv} ,
\label{eq:coeff_conv}
\end{equation}
where $\ell_{\rm conv} = \alpha H_P$ is the mixing length (with $H_P$ denoting the pressure scale height) and $v_{\rm conv}$ is the convective velocity obtained from the mixing-length theory (see Appendix \ref{sec:app_a_conv}). Just outside a convection zone, mixing due to convective overshoot is included using an exponentially decaying diffusion coefficient, as suggested by \citet{freytag1996},
\begin{equation}
D_{\rm ov} = D_{{\rm conv},0} \ \exp{ \left( \frac{-2 |r-r_0|}{f_{\rm ov} \, H_{P,0}} \right) } ,
\end{equation}
where $r_0$ is the radial boundary of the convection zone, $H_{P,0}$ and $D_{{\rm conv},0}$ are the pressure scale height and convective mixing coefficient at that radius, and $f_{\rm ov}$ is a free parameter controlling the extent of the overshooting region.

STELUM also has the capability to model mixing processes arising in the presence of composition gradients. On one hand, the phenomenon of semi-convection occurs in a region that is convectively unstable according to the Schwarzschild criterion but where the composition gradient has a stabilizing effect. Following \citet{langer1985}, the semi-convective diffusion coefficient takes the form
\begin{equation}
D_{\rm sc} = \frac{8}{9} \alpha_{\rm sc} \frac{\sigma T^3}{\rho^2 \kappa c_P} \frac{ \nabla_{\rm rad} - \nabla_{\rm ad} }{ B - \left( \nabla_{\rm rad} - \nabla_{\rm ad} \right) } ,
\label{eq:coeff_sc}
\end{equation}
where $c_P$ is the specific heat capacity at constant pressure, $B$ is the so-called Ledoux term (which measures the influence of the composition gradient), and $\alpha_{\rm sc}$ is an adjustable efficiency factor. On the other hand, the reverse situation, a stable thermal structure combined with an unstable chemical structure, leads to the process of thermohaline convection (also called fingering convection). Based on the work of \citet{kippenhahn1980}, the thermohaline mixing coefficient can be expressed as
\begin{equation}
D_{\rm th} =  8 \alpha_{\rm th} \frac{\sigma T^3}{\rho^2 \kappa c_P} \frac{B}{ \nabla_{\rm rad} - \nabla_{\rm ad} } ,
\label{eq:coeff_th}
\end{equation}
where $\alpha_{\rm th}$ is another efficiency parameter\footnote{Note that Equations \ref{eq:coeff_sc} and \ref{eq:coeff_th} assume that the actual temperature gradient is given by the radiative temperature gradient, which means that we ignore the energy transport associated with semi-convective and thermohaline mixing.}. We rely on standard quantitative criteria to determine the onset of semi-convective and thermohaline mixing (see for instance \citealt{salaris2017}).

Like the structure equations, the $I$ transport Equations \ref{eq:trans} are first recast in terms of the independent mass variable $\xi$ defined in Equation \ref{eq:xi_to_q}. The numerical solution is then performed simultaneously for all atomic species using the implicit Runge-Kutta integrator RADAU5 of \citet{hairer1996}, which is particularly well suited for stiff problems. In the present case, the stiffness of the problem is due to the large disparity of diffusive timescales between the center and surface of a star and between the various transport mechanisms. For numerical convenience, our chemical transport formalism is not applied all the way up to the surface, where the diffusive timescales can become extremely short, but only up to a pre-specified fractional mass limit $q_{\rm lim}$. Above this layer ($q < q_{\rm lim}$), two choices of boundary condition are possible. The first option is simply to assume a uniform composition (that is, the composition at $q = q_{\rm lim}$ is imposed everywhere above). The second option is to solve a simplified form of the transport equation inspired from \citet{charbonneau1993} and \citet{turcotte1993},
\begin{equation}
\frac{dX_i}{dt} = S_{{\rm wind},i} + S_{{\rm acc},i} + \frac{1}{r^2 \rho} \frac{d}{dr} \left( r^2 \rho D_{\rm lim} \frac{dX_i}{dr} \right) ,
\end{equation}
where $S_{{\rm wind},i}$ and $S_{{\rm acc},i}$ are source/sink terms accounting for mass loss and accretion, and $D_{\rm lim}$ is a pre-specified diffusion coefficient designed to generate artificial mixing. This method is computationally more expensive but more stable when mass loss or accretion is present. The source/sink terms are given by
\begin{align}
S_{{\rm wind},i} &= \frac{\dot{M}_{\rm wind}}{q_{\rm lim} M} X_i , \\[5pt]
S_{{\rm acc},i} &= \frac{\dot{M}_{\rm acc}}{q_{\rm lim} M} \left( X_{{\rm acc},i} - X_i \right) ,
\end{align}
where $X_{{\rm acc},i}$ is the mass fraction of element $i$ in the accreted material.

\section{Illustrative Simulations} \label{sec:simu}

The aim of this section is to demonstrate the modeling capabilities of STELUM in the framework of the spectral evolution of white dwarfs. To do so, we present two illustrative simulations of this phenomenon. In the first case, we follow the transformation of a helium-, carbon-, and oxygen-rich PG 1159 star into a helium-atmosphere DB white dwarf due to the downward diffusion of heavy elements, and then into a helium-dominated, carbon-polluted DQ white dwarf through the convective dredge-up of settling carbon. This is often called the PG 1159$-$DO$-$DB$-$DQ evolutionary channel. In the second case, we compute the evolution of a hot helium-rich DO white dwarf containing a small amount of hydrogen, which becomes a hydrogen-atmosphere DA star as a consequence of the float-up of hydrogen, and then a helium-atmosphere DC star through the convective mixing of the hydrogen and helium layers. We refer to this as the DO$-$DA$-$DC evolutionary channel. As discussed in the Introduction, there is strong empirical evidence that both scenarios indeed occur in nature.

Our adopted computational setup is as follows. For both simulations, we consider a typical mass $M = 0.6$ $\msun$ and a very high initial effective temperature $\Teff = 90,000$ K. The chemical structure consists of a homogeneous, equimassic carbon/oxygen core and a homogeneous envelope of fractional mass $\log q_{\rm env} = -2.0$. The initial envelope composition is the only parameter that differs between the two simulations and is specified below. The following element transport mechanisms are taken into account: chemical, gravitational, and thermal diffusion (including non-ideal effects), convective and overshoot mixing, and an outward wind in the early evolution. We use mixing-length and overshoot parameters of $\alpha = 1.0$ and $f_{\rm ov} = 0.075$; the reason for this choice will become apparent below. We assume the wind mass-loss law of \citet{blocker1995},
\begin{equation}
\dot{M}_{\rm wind} = -1.29 \times 10^{-15} \left( \frac{L}{\lsun} \right)^{1.86} \msun/\rm{yr} ,
\label{eq:rate_wind}
\end{equation}
where $L$ is the total luminosity of the star. This prescription not only has the advantage of computational simplicity, but it also has proven appropriate for hot white dwarfs, even though it was originally developed to model post-AGB stars (see below). In order to capture the chemical evolution of the atmosphere, the transport boundary is placed very high in the envelope, at fractional mass $\log q_{\rm lim} = -14.0$. We assume a uniform composition above this point. Finally, residual nuclear burning is ignored throughout the cooling process.

\subsection{The PG 1159$-$DO$-$DB$-$DQ Evolutionary Channel} \label{sec:simu_DO-DB-DQ}

The chain of events linking the PG 1159 stars and DQ white dwarfs is perhaps the best-studied case of spectral evolution and thus constitutes an important benchmark for our demonstration of the capabilities of STELUM. Theoretical calculations of the PG 1159-to-DO transition were performed by a number of authors over the years, namely, \citet{dehner1995}, \citet{unglaub2000}, \citet{fontaine2002}, \citet{althaus2004}, and \citet{quirion2012}. The DB-to-DQ transition has also been the focus of detailed modeling, starting with the work of \citet{pelletier1986} and \citet{macdonald1998}. Later on, simulations of the entire PG 1159$-$DO$-$DB$-$DQ evolution were carried out by the La Plata group in \citet{althaus2005a}, \citet{scoccola2006}, and \citet{camisassa2017}, and by our own group in \citet{dufour2005} using the ancestor of STELUM (see also \citealt{fontaine2005} and \citealt{brassard2007}). Obviously, these various sets of models rely on disparate physical and numerical assumptions, especially regarding element transport. An exhaustive discussion of these differences is beyond the scope of the present paper and is thus deferred to a future publication (A. B\'edard et al., in preparation). For now, we simply describe the results of our new calculations and make brief qualitative comparisons with previous works wherever appropriate.

\begin{figure*}
\centering
\includegraphics[width=2.\columnwidth,clip=true,trim=2.0cm 5.0cm 1.5cm 6.5cm]{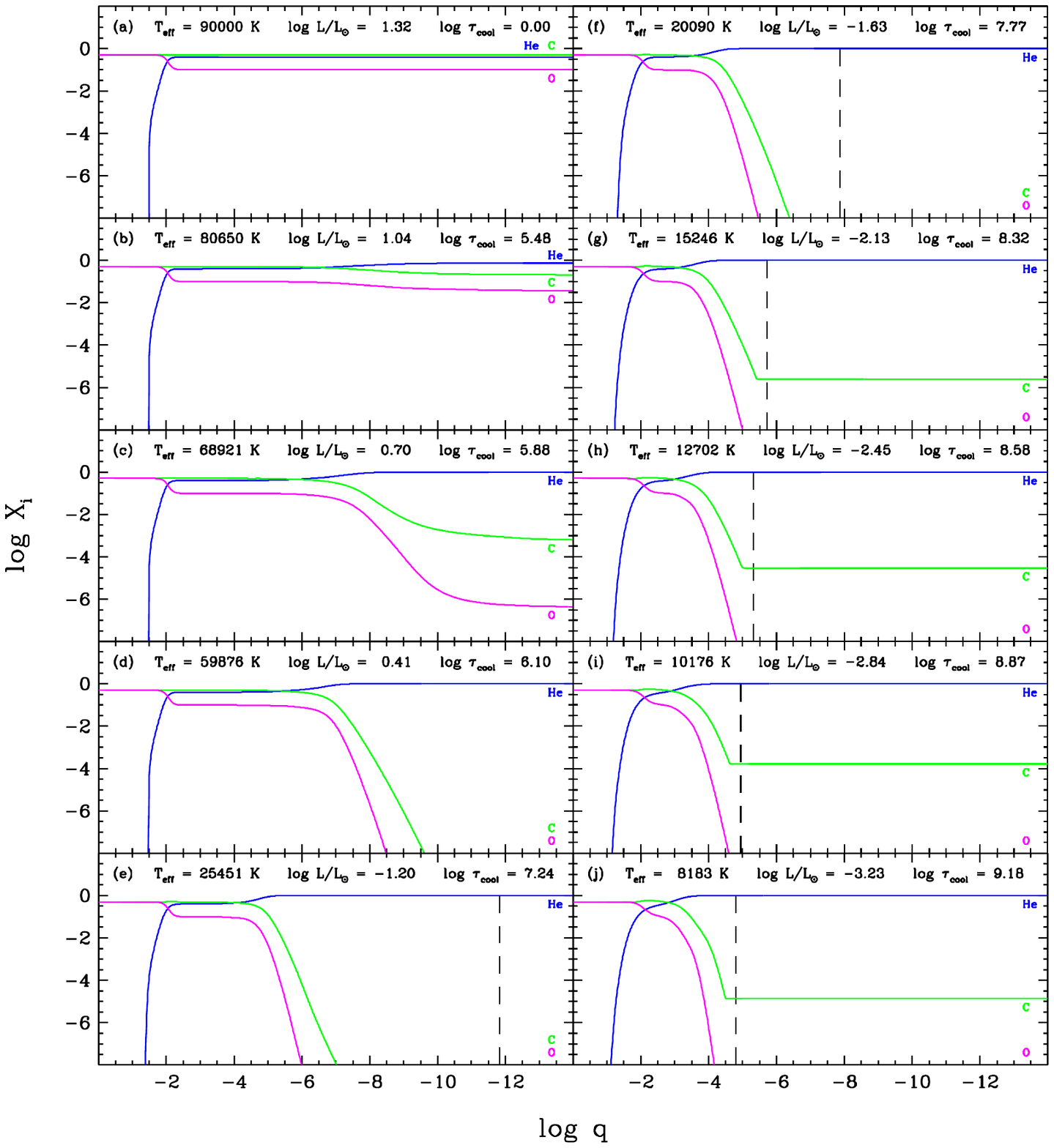}
\caption{Snapshots of the chemical structure, represented here as the run of the elemental mass fractions with fractional mass depth ($q = 1 - m/M$), at various points along our PG 1159$-$DO$-$DB$-$DQ evolutionary simulation. The hydrogen, helium, carbon, and oxygen abundance profiles are shown as red, blue, green, and magenta curves, respectively. The location of the base of the convection zone is indicated by a black dashed line. Each panel also gives the effective temperature, total luminosity, and cooling age of the displayed model. A video showing the full simulation is available as supplementary material in the online version of this article (see Appendix \ref{sec:app_b}).}
\vspace{2mm}
\label{fig:evol_DO-DB-DQ}
\end{figure*}

In our simulation, the initial envelope composition is a helium/carbon/oxygen mixture typical of PG 1159 stars: $X_{\rm He} = 0.4$, $X_{\rm C} = 0.5$, and $X_{\rm O} = 0.1$ \citep{werner2006}. Figure \ref{fig:evol_DO-DB-DQ} shows the chemical profile (the elemental mass fractions $\log X_i$ as a function of fractional mass depth $\log q$) at various key points along the evolutionary sequence. The initial model is displayed in panel (a). Note that this starting chemical profile is approximate; the only way to obtain more realistic initial conditions is to compute the full evolution prior to the white dwarf stage, including the late helium-shell flash and associated born-again episode, as done for instance in \citet{althaus2005a}. A video showing the evolving chemical structure is available online as supplementary material (see Appendix \ref{sec:app_b}).

The early chemical evolution is dominated by the stellar wind, which is strong enough to counteract gravitational settling. Therefore, the carbon and oxygen are temporarily maintained in the outer layers, and the star still exhibits a PG 1159-like surface composition at $\Teff \sim 80,000$ K, as can be seen in panel (b). However, given the adopted expression for the mass-loss rate, the cooling of the star causes the wind to fade and thus to lose its ability to compete against gravitational settling. As a result, below $\Teff \sim 75,000$ K, the superficial carbon and oxygen begin to sink rapidly into deeper layers, thereby producing a helium-atmosphere DO white dwarf. After a short transient phase in which a trace amount of carbon is still detectable, the outer envelope becomes completely devoid of heavy elements below $\Teff \sim 60,000$ K, as displayed in panels (c) and (d), respectively. 

It is worth emphasizing that the choice to include a wind in our calculations is far from arbitrary. In an analogous simulation without mass loss (not shown here), the carbon and oxygen diffuse out of the atmosphere in a few years, such that the PG 1159-to-DO transition occurs essentially instantaneously at the onset of the evolution. Such a behavior is clearly at odds with the existence of PG 1159 stars over a wide range of effective temperatures \citep{werner2006}. This observational evidence necessarily implies that some mechanism, most likely a wind, prevents gravitational settling at the beginning of the cooling sequence. Previous theoretical investigations by \citet{unglaub2000} and \citet{quirion2012} reached a similar conclusion. Moreover, \citet{quirion2012} examined the effect of changing the mass-loss prescription and found that Equation \ref{eq:rate_wind} provides the best match to the observed red edge of both the PG 1159 spectroscopic domain and the GW Vir instability strip, hence our use of this particular formula. Interestingly, we also note that our simple wind model predicts $\log X_{\rm C} \sim -3.0$ at the surface around $\Teff \sim 70,000$ K, which is broadly consistent with measured carbon abundances of hot DO white dwarfs \citep{dreizler1996,reindl2014}.

During the subsequent evolution, the carbon and oxygen continue to sink further into the star under the influence of gravitational settling and thermal diffusion. The pure-helium layer consequently grows from $\log q_{\rm He} \sim -8.0$ at $\Teff \sim 60,000$ K to $\log q_{\rm He} \sim -6.0$ at $\Teff \sim 25,000$ K, as shown in panel (e). However, in this DB white dwarf, element separation is still far from complete: the envelope harbors a double-layered chemical structure, with a pure-helium layer on top of a mantle retaining the initial composition of the PG 1159 progenitor. This is due to the fact that the diffusion timescales are much longer at the bottom than at the top of the envelope. This result was first demonstrated by \citet{dehner1995} and later corroborated by \citet{fontaine2002} and \citet{althaus2004}. As discussed below, it is of fundamental importance for our understanding of DQ white dwarfs. Besides, note that the existence of the double-layered structure is supported independently by asteroseismic analyses of pulsating DB stars \citep{giammichele2018,bischoff-kim2019}. In particular, the envelope composition profile displayed in panel (e) closely resembles that of the prototypical DB pulsator GD 358 as inferred from asteroseismology.

At this stage of the evolution, our DB white dwarf has a superficial convection zone associated with the recombination of helium, which grows deeper with further cooling \citep{rolland2018,cukanovaite2019}. The location of the base of the convective region is shown by a black dashed line in Figure \ref{fig:evol_DO-DB-DQ}. Comparing panels (e) and (f), it can be seen that the deepening of the convection zone occurs much faster than the sinking of carbon and oxygen: between $\Teff \sim 25,000$ and 20,000 K, the mass extent of the convective region increases by four orders of magnitude, while the chemical profile barely changes. The inevitable outcome is that the helium convection zone eventually catches up some of the settling carbon, which is thus brought back to the surface by virtue of the highly efficient convective mixing, as shown in panel (g). Notice that the region in which carbon is uniformly mixed (corresponding to the flat part of the carbon abundance profile) actually extends slightly beyond the formal convective boundary as a result of convective overshoot. As the star cools and the convective motions reach deeper, more carbon-rich layers, the amount of carbon in the helium-rich envelope increases. This is clearly seen in panels (g), (h), and (i): the surface mass fraction is $\log X_{\rm C} \sim -5.6$, $-4.5$, and $-3.8$ at $\Teff \sim 15,200$, 12,700, and 10,200 K, respectively. The carbon becomes spectroscopically visible, hence making the white dwarf a member of the DQ spectral class. 

Then, at even lower effective temperatures, the trend reverses: after a plateau between $\Teff \sim 10,600$ and 9800 K, the surface carbon abundance starts decreasing, as shown in panel (j). This behavior originates from a combination of two phenomena. First, the base of the convection zone reaches its maximum depth and thereby stops carrying additional carbon to the surface. Second, as originally explained in detail by \citet{pelletier1986}, a change in the average ionization state of carbon makes the slope of the carbon diffusion tail steeper, with the result that carbon partially sinks out of the convective region. This effect is apparent when comparing panels (i) and (j). 

\begin{figure}
\centering
\includegraphics[width=\columnwidth,clip=true,trim=2.5cm 4.75cm 2.0cm 6.25cm]{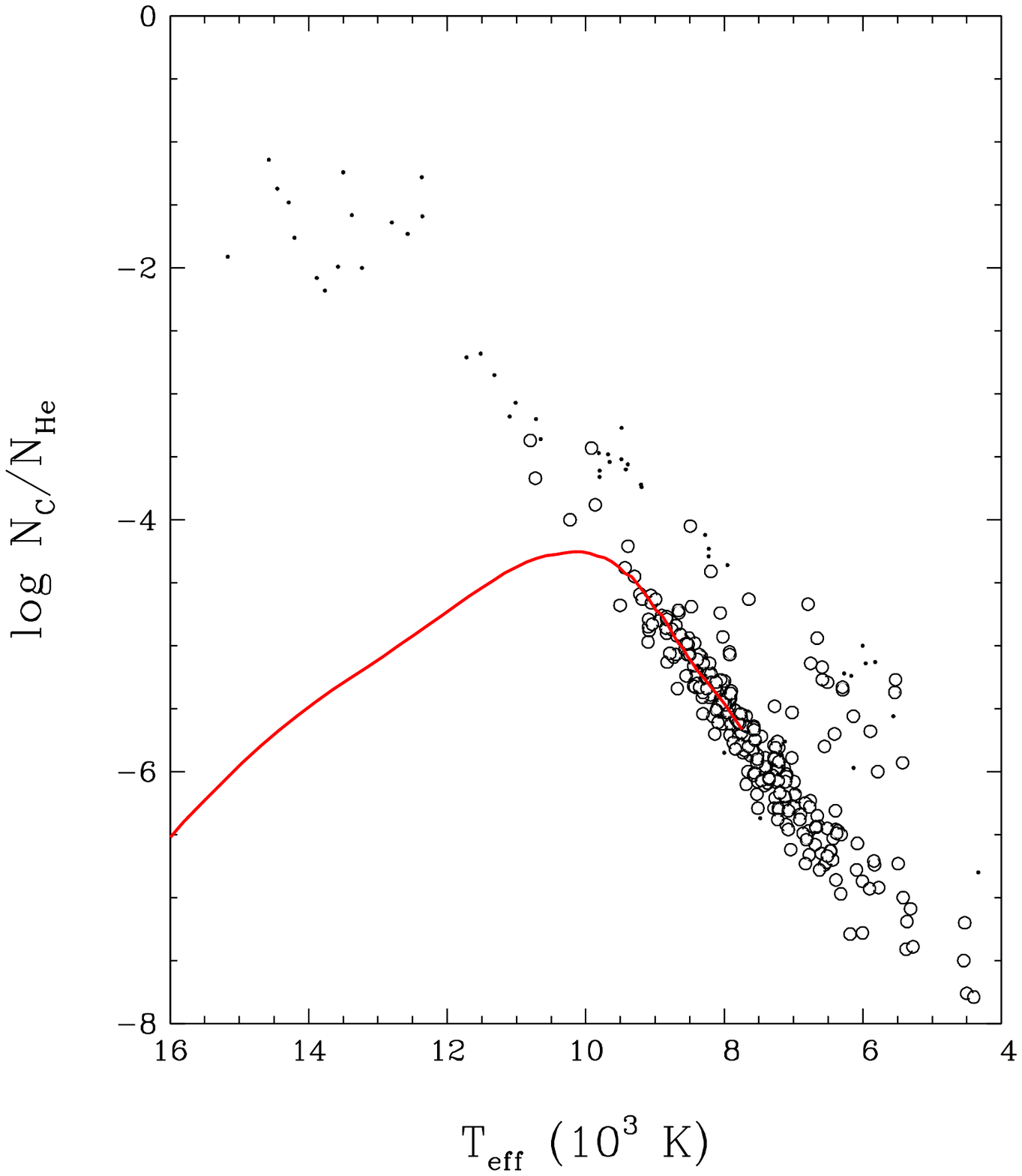}
\caption{Theoretical and empirical relations between the atmospheric carbon-to-helium number ratio and the effective temperature of DQ white dwarfs. The red curve shows the prediction from our PG 1159$-$DO$-$DB$-$DQ evolutionary simulation, while the black symbols represent empirical atmospheric parameters of DQ stars taken from \citet{coutu2019} and \citet{blouin2019b}. The objects with $M \le 0.7$ $\msun$ and $M > 0.7$ $\msun$ are displayed as large circles and small dots, respectively.}
\vspace{2mm}
\label{fig:compseq_DQ}
\end{figure}

The relation between surface carbon abundance and effective temperature predicted by our chemical evolution simulation can be compared to empirically determined atmospheric parameters of DQ stars. Figure \ref{fig:compseq_DQ} shows our model prediction as a red curve together with DQ white dwarfs from \citet{coutu2019} and \citet{blouin2019b} as black symbols in the $\log N_{\rm C}/N_{\rm He}-\Teff$ diagram (where $N_{\rm C}/N_{\rm He}$ denotes the atmospheric carbon-to-helium number ratio). In the following, we focus solely on the classical, normal-mass DQ stars ($M \le 0.7$ $\msun$; large circles) and refrain from discussing the more massive objects ($M > 0.7$ $\msun$; small dots). The latter group is believed to be produced by white dwarf mergers and thus constitutes a fundamentally different population \citep{dunlap2015,coutu2019,cheng2019}, so our calculations obviously do not apply to these objects. Most classical DQ stars form a tight sequence in the $\log N_{\rm C}/N_{\rm He}-\Teff$ diagram, following a clear trend of decreasing carbon abundance with decreasing temperature. Figure \ref{fig:compseq_DQ} demonstrates that our simulation perfectly reproduces both the location and the slope of the observed relation. At this time, our calculations cover only the first half of the DQ sequence and cannot be extended to lower effective temperatures due to the unavailability of OPAL radiative opacities for carbon in this region of parameter space. Besides, the absence of objects on the ascending part of the theoretical curve is actually a selection effect: in this temperature range, white dwarfs retain their DB spectral type because the carbon abundance is well below the optical detection limit. Nevertheless, ultraviolet observations of a few cool DB stars have revealed the expected amount of atmospheric carbon \citep{desharnais2008}.

It is important to mention that the agreement between the data and our model is not entirely coincidental. In fact, the height of the predicted curve in the $\log N_{\rm C}/N_{\rm He}-\Teff$ diagram depends very sensitively on a number of model parameters, most notably the stellar mass, the depth of the envelope, the initial carbon abundance of the PG 1159 progenitor, and the extent of convective overshoot. For standard values of the first three parameters ($M = 0.6$ $\msun$, $\log q_{\rm env} = -2.0$, $X_{\rm C} = 0.5$), the overshoot efficiency factor $f_{\rm ov}$ can be adjusted to match the location of the observed DQ sequence. We find that $f_{\rm ov} = 0.075$ yields the good agreement shown in Figure \ref{fig:compseq_DQ}, hence our choice of this specific value. Interestingly, 3D hydrodynamical simulations of convection in white dwarfs favor a higher overshoot parameter, $f_{\rm ov} \sim 0.2-0.4$ (\citealt{cunningham2019}; T. Cunningham 2021, private communication). However, these calculations are restricted to shallow convection zones extending no deeper than $\log q \sim -12.0$. Given that DQ stars have deep convection zones reaching down to $\log q \sim -5.0$, our results indicate that the efficiency of convective overshoot is strongly depth-dependent, a conclusion also reached in the context of AGB stars \citep{herwig2000}. Therefore, our simulations could potentially be used to calibrate the extent of overshoot mixing in cool white dwarfs, for which 3D hydrodynamical models are currently unavailable. An exhaustive exploration of the model parameter space, which is obviously beyond the scope of the present work, will be the focus of a forthcoming publication (A. B\'edard et al., in preparation).

Regardless of this matter, we want to stress that it is still remarkable that our evolutionary sequence successfully reproduces the slope of the empirical $\log N_{\rm C}/N_{\rm He}-\Teff$ relation at low temperatures. Indeed, as noted earlier, the dredge-up of carbon in helium-rich white dwarfs and the corresponding DB-to-DQ transition have been the subject of numerous theoretical studies, both by our group \citep{pelletier1986,fontaine2005,dufour2005,brassard2007} and by other groups \citep{macdonald1998,althaus2005a,scoccola2006,camisassa2017}. Among these, only the models of our group have been able to match the observed trend of decreasing carbon abundance with decreasing effective temperature, while other models predict instead a constant or increasing carbon abundance. The differences between previous works and our present calculations will be further discussed in our upcoming paper.

Finally, one last clarification needs to be made. The first generation of DQ models developed in the 1980s assumed chemical structures consisting of a single-layered, helium-dominated envelope atop a pure-carbon core in diffusive equilibrium. In this framework, the trace carbon at the surface is dredged-up from the core, a process which can only take place in envelopes much thinner ($\log q_{\rm env} \sim -3.5$) than expected from the standard theory of stellar evolution \citep{koester1982,fontaine1984,pelletier1986}. We have known since the work of \citet{dehner1995} that this scenario contains a major flaw: the envelope of a DQ star has not had the time to achieve complete element separation and is therefore characterized by the double-layered chemical profile discussed above. Consequently, the atmospheric carbon does not come from the core, but rather from the PG 1159-like layer at the bottom of the envelope, as clearly seen in Figure \ref{fig:evol_DO-DB-DQ}. This distinction is crucial: the modern paradigm requires more standard envelope masses ($\log q_{\rm env} \sim -2.0$), thereby resolving the conflict between stellar evolution theory and the existence of carbon-polluted white dwarfs. Unfortunately, the misconception that DQ stars have reached diffusive equilibrium and can thus only be explained by thin envelopes is still widespread in the current literature \citep{koester2020}.

\subsection{The DO$-$DA$-$DC Evolutionary Channel} \label{sec:simu_DO-DA-DC}

The DO$-$DA$-$DC connection represents another frequently invoked case of white dwarf spectral evolution. Indeed, the DO-to-DA and DA-to-DC transitions have been the topic of many theoretical investigations, although they have often been examined and discussed separately, as two unrelated phenomena. The float-up of residual hydrogen at high effective temperature was first modeled by \citet{unglaub1998,unglaub2000}. Such calculations were also performed by \citet{althaus2005b,althaus2020}, who however placed little emphasis on the chemical transformation itself and focused instead on its implications for stellar pulsations. The convective mixing of the hydrogen and helium layers at low effective temperature has an even longer modeling history, comprising the works of \citet{baglin1973}, \citet{koester1976}, \citet{dantona1989}, \citet{macdonald1991}, \citet{althaus1998}, \citet{chen2011}, and \citet{rolland2018}. It is beyond the scope of the present paper to review all of the aforementioned studies and their very diverse levels of sophistication. Nevertheless, while describing our own simulation below, we briefly comment on its similarities and differences with older calculations. For now, we want to point out that this is the first time that the DO$-$DA$-$DC evolution is computed as a whole, from beginning to end, including both the float-up process at high temperature and the convective mixing event at low temperature.

\begin{figure*}
\centering
\includegraphics[width=2.\columnwidth,clip=true,trim=2.0cm 3.25cm 1.5cm 5.25cm]{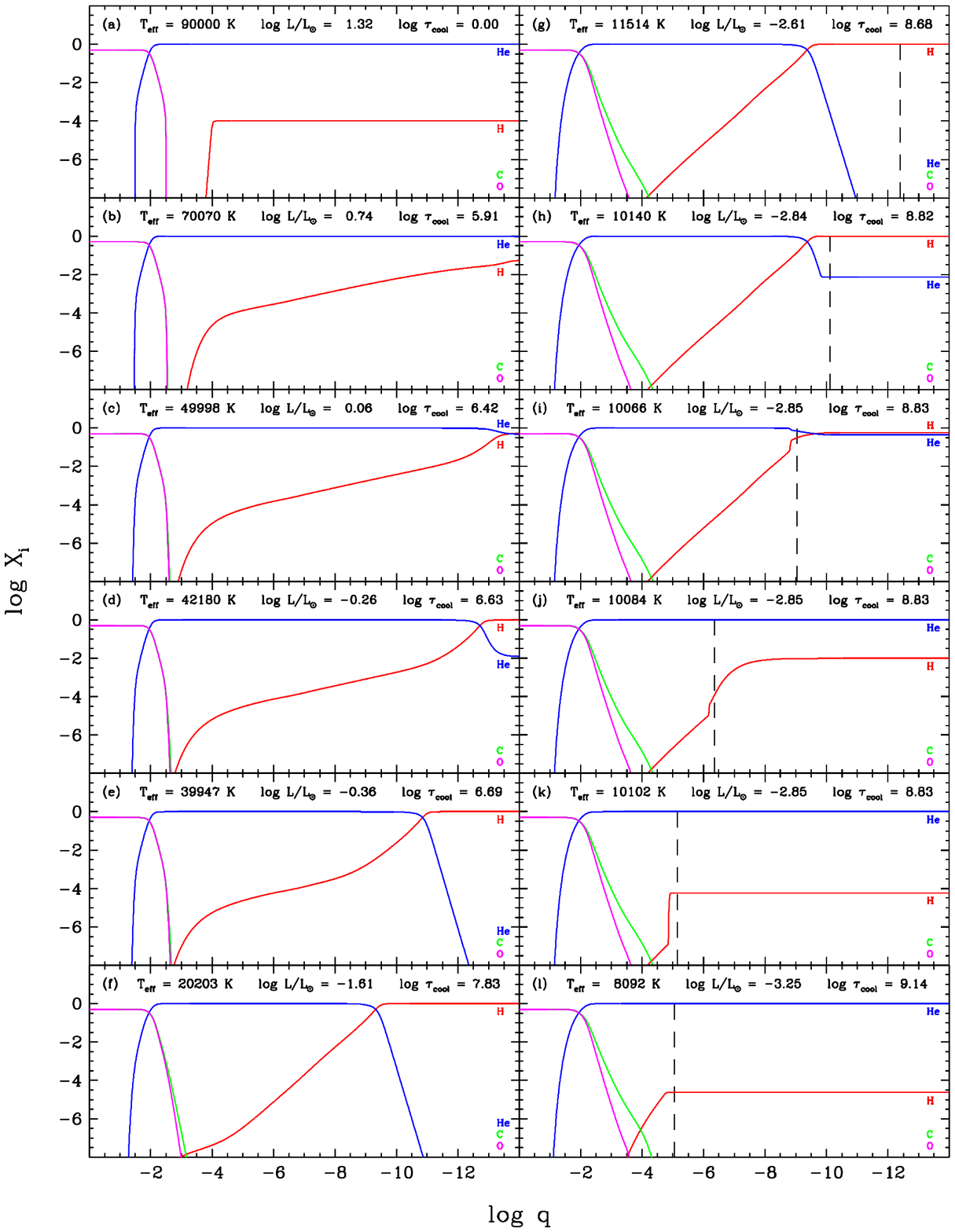}
\caption{Same as Figure \ref{fig:evol_DO-DB-DQ}, but for our DO$-$DA$-$DC evolutionary simulation.}
\vspace{2mm}
\label{fig:evol_DO-DA-DC}
\end{figure*}

Our simulation begins with a helium-dominated envelope containing a small amount of hydrogen: $X_{\rm H} = 0.0001$, $X_{\rm He} = 0.9999$. This initial model would be classified as a normal DO star, given that the optical detection limit of hydrogen in a hot helium-rich atmosphere is $X_{\rm H} \sim 0.01$ \citep{werner1996}. Apart from the initial envelope composition, we use the same computational settings as in our previous simulation. Figure \ref{fig:evol_DO-DA-DC} details the evolution of the chemical structure, with the starting model shown in panel (a). An important thing to note is that the trace hydrogen profile does not extend down to the base of the envelope, but only down to $\log q = -4.0$, because deeper-lying hydrogen is expected to be burned in previous evolutionary phases \citep{althaus2005b}. Consequently, the total hydrogen mass is 10$^{-8} M$. Once again, a video displaying the full chemical evolution is available online as supplementary material (see Appendix \ref{sec:app_b}).

Panels (a) to (f) outline the float-up of hydrogen and the associated DO-to-DA transformation taking place in the early cooling phase. At first, the float-up process is considerably slow, because the outer hydrogen-enriched layers are successively peeled off by the stellar wind. Consequently, the hydrogen is still spectroscopically invisible at $\Teff \sim 70,000$ K. The surface eventually becomes richer in hydrogen than in helium around $\Teff \sim 50,000$ K. Thereafter, the superficial helium quite rapidly sinks out of sight, leaving a pure-hydrogen atmosphere at $\Teff \sim 40,000$ K. The hydrogen located deeper in the envelope continues to diffuse upward, such that the pure-hydrogen layer gradually becomes thicker. At $\Teff \sim 20,000$ K, element separation is complete: the hydrogen and helium layers have reached diffusive equilibrium. Note that at this point, the total hydrogen mass is now $\sim$10$^{-9} M$, meaning that $\sim$90\% of the hydrogen initially present in the model has been ejected by the wind. The leftovers form a pure-hydrogen layer of thickness $\log q_{\rm H} \sim -10.0$ and an extended diffusion tail underneath.

Interestingly, our simulated DO-to-DA transition occurs at the right time in the evolution, that is, in an effective temperature range where the fraction of DA stars is observed to increase (Paper I). Furthermore, our calculation shows that this transformation takes place on a Myr timescale, and thus that there is a non-negligible probability to observe objects currently transitioning from a helium atmosphere to a hydrogen atmosphere. This is entirely consistent with the recent detection of tens of hot white dwarfs possessing extremely thin hydrogen layers (\citealt{manseau2016}; Paper I). In that respect, we must stress once again the important role played by the wind: without the effect of mass loss, gravitational settling would be so efficient that all DO stars containing residual hydrogen would become DA stars practically instantaneously. This scenario is in stark conflict with our empirical knowledge of the spectral evolution of white dwarfs, thereby strongly supporting the existence of stellar winds at the beginning of the cooling sequence. Qualitatively similar results were obtained by \citet{unglaub1998,unglaub2000} from transport calculations in static envelope models.

The immediately subsequent evolution is relatively uneventful: since the envelope is in diffusive equilibrium, the chemical profile remains essentially unchanged, as illustrated by the model with $\Teff \sim 11,500$ K in panel (g). This model falls within the ZZ Ceti instability strip and is therefore representative of a pulsating DA white dwarf with a thin hydrogen layer, such as Ross 548 \citep{giammichele2016}. 

However, this is the calm before the storm. From that point on, the recombination of hydrogen produces a superficial convection zone, which grows deeper with further cooling \citep{tremblay2015}. The black dashed line in Figure \ref{fig:evol_DO-DA-DC} indicates the location of the base of the convective region. A radical chemical transformation follows: the advective motions reach into the underlying helium, which is thus dredged-up and thoroughly mixed into the outer hydrogen layer, as can be seen in panel (h). This phenomenon is closely analogous to the dredge-up of carbon that turns a DB star into a DQ star, as featured in our previous simulation. The main difference here is that we are dealing with a runaway process. At a given effective temperature, a helium-rich convection zone is much deeper than a hydrogen-rich convection zone \citep{rolland2018}. Consequently, the increase in helium abundance results in a more extended convective region, which in turn brings more helium in the outer layers, and so on. In the end, the hydrogen previously accumulated at the surface is entirely diluted in the helium-dominated envelope. This phenomenon is well illustrated in panels (h) to (k): as the base of the convection zone moves inward from $\log q \sim -10.0$ to $-5.0$, hydrogen goes from dominating the atmospheric composition to becoming a trace element.

A few aspects of this mixing episode are particularly worthy of interest and deserve further comments. First, the chemical transformation occurs extremely rapidly, as indicated by the constant cooling age given in panels (i) to (k). In fact, the timescale for the growth of the convection zone is shorter than the mixing and diffusion timescales. This is obvious in panel (j), where the non-flat hydrogen abundance profile reveals that the convective region has not had the time to achieve complete mixing, and in panel (k), where the steep hydrogen abundance gradient below the convection zone has not yet been smoothed out by chemical diffusion. Second, the mixing event causes a slight increase in effective temperature, from $\Teff = 10,062$ to 10,102 K, as partly shown in panels (i) to (k). This behavior, highly unusual by white dwarf standards, was also reported in a few earlier studies \citep{baglin1973,dantona1989,chen2011}. It was explained by \citet{chen2011} as the result of the opacity change in the outer layers and the convective coupling between the core and the envelope. We note in passing that all previous theoretical calculations of convective mixing \citep{baglin1973,koester1976,dantona1989,macdonald1991,althaus1998,chen2011,rolland2018} relied on simplified semi-evolutionary approaches and/or instantaneous mixing approximations. To our knowledge, ours is the first fully self-consistent simulation of this phenomenon, including both a detailed time-dependent treatment of element transport and the feedback of composition changes on the evolution of the star.

Once the convection zone stabilizes, the cooling resumes on a more typical timescale, thereby allowing the hydrogen abundance profile below the convection zone to relax toward diffusive equilibrium. The chemical structure of the model with $\Teff \sim 8100$ K is displayed in panel (l). The final surface hydrogen mass fraction is $\log X_{\rm H} \sim -4.6$, just below the optical detection limit in this effective temperature range \citep{rolland2018}. Our DA star has therefore become a DC star. Besides, note that this conversion necessarily involves a short phase during which our model would appear as a so-called helium-rich DA white dwarf.

\begin{figure}
\centering
\includegraphics[width=\columnwidth,clip=true,trim=2.5cm 4.75cm 2.0cm 6.25cm]{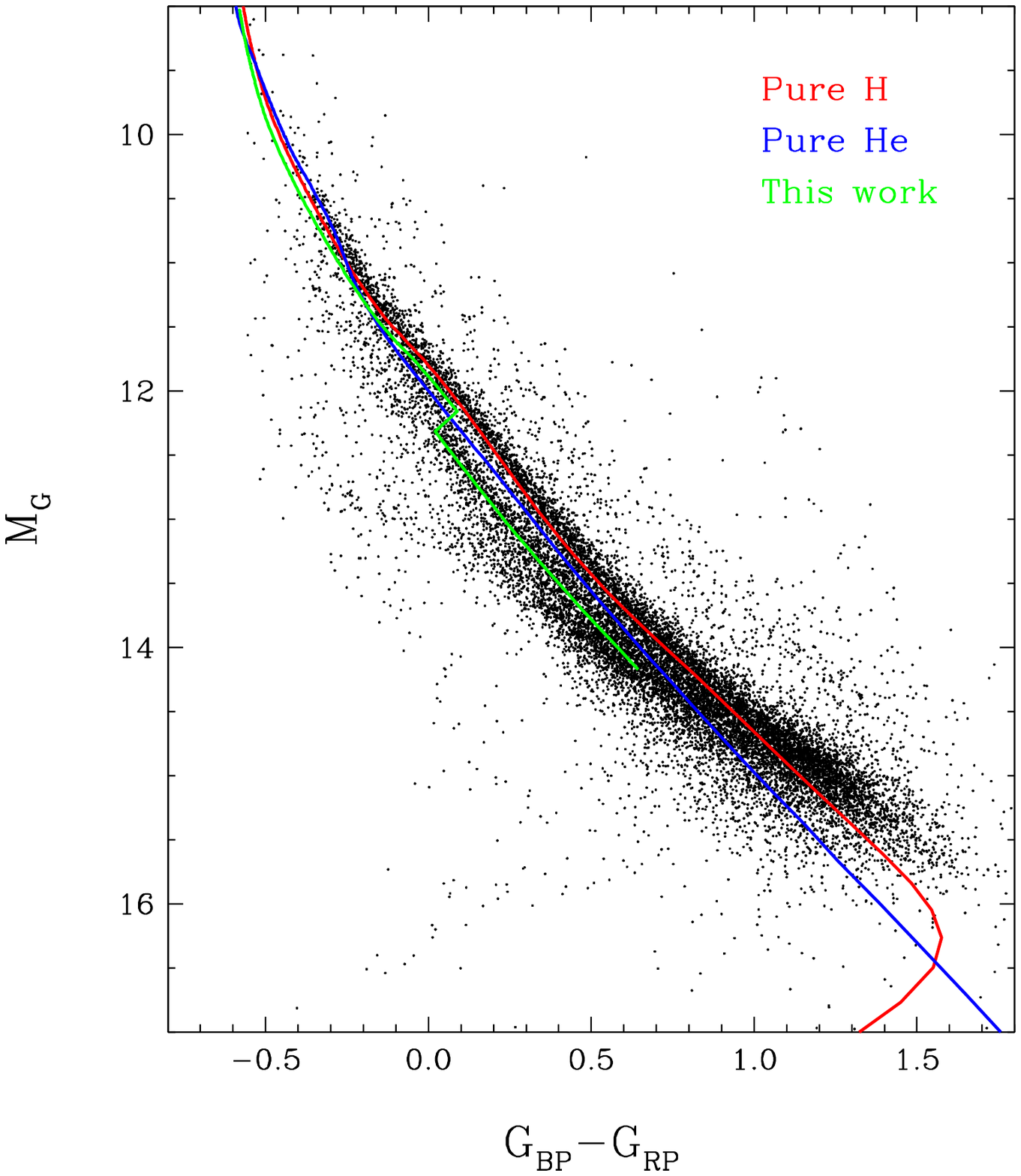}
\caption{Color-magnitude diagram showing the Gaia sample of white dwarfs within 100 pc taken from the MWDD, together with theoretical cooling tracks of 0.6 $\msun$ white dwarfs with various atmospheric compositions: pure hydrogen, pure helium, and the hydrogen-helium mixture predicted by our DO$-$DA$-$DC evolutionary simulation, which correspond to the red, blue, and green curves, respectively.}
\vspace{2mm}
\label{fig:compseq_DC}
\end{figure}

A few pieces of empirical evidence firmly support the occurrence of such a DA-to-DC transition in real stars. First, this phenomenon is believed to be responsible for the observed increase of the fraction of helium-atmosphere white dwarfs at low effective temperatures \citep{blouin2019a,cunningham2020}. Second, it also provides an elegant explanation to the famous bifurcation seen in the Gaia white dwarf sequence \citep{gaia2018}. This is illustrated in Figure \ref{fig:compseq_DC}, which shows the color-magnitude diagram of the Gaia sample of white dwarfs within 100 pc \citep{gaia2018} taken from the Montreal White Dwarf Database (MWDD; \citealt{dufour2017}). Also displayed are theoretical cooling tracks of 0.6 $\msun$ white dwarfs with various atmospheric compositions (\citealt{holberg2006,tremblay2011,bergeron2011,blouin2018}; Paper I). The red and blue curves correspond to pure-hydrogen and pure-helium atmospheres, respectively, while the green curve assumes the varying surface composition predicted by our simulation. As noted in many previous works, the pure-hydrogen sequence nicely coincides with the upper branch of the bifurcation, while the pure-helium sequence fails to reproduce the lower branch of the bifurcation. It has been demonstrated that the atmosphere of the objects populating the latter region is indeed helium-rich, but also likely contains an undetectable amount of hydrogen (or any other electron donor) which is the source of the color and magnitude offset \citep{bergeron2019}. Figure \ref{fig:compseq_DC} shows that our simulation is in excellent agreement with this interpretation. The green curve initially follows the red curve and heads toward the upper branch, as expected for a DA star, but then abruptly shifts downward and subsequently moves along the lower branch as a result of convective mixing. Both the location and magnitude of the predicted shift are remarkably consistent with the morphology of the observed color-magnitude diagram, indicating that the mixing temperature and hydrogen abundance of our final DC model are realistic. In short, our results confirm that the lower branch of the bifurcation can be ascribed to helium-dominated atmospheres with traces of hydrogen and thus constitutes an observational signature of convectively driven spectral evolution.

Of course, the evolutionary calculation presented here involves several physical parameters whose influence has yet to be examined quantitatively. The most important variable is certainly the mass fraction of hydrogen in the initial model. A larger amount of hydrogen is expected to result in an earlier DO-to-DA transition, a later DA-to-DC transition, and a larger final hydrogen abundance. Another critical model parameter is the mass-loss law, which dictates the strength of the wind and hence the efficiency of gravitational settling at high effective temperatures. While the expression adopted here (Equation \ref{eq:rate_wind}) was shown to be appropriate for PG 1159 stars, it likely overestimates the mass-loss rate in DO white dwarfs poorer in heavy elements, given that the wind is thought to be metal-driven. The use of a more realistic mass-loss law might make the float-up of hydrogen faster. Furthermore, at lower effective temperatures, the convective mixing of the hydrogen and helium layers obviously depends on the extent of convective overshoot. We assumed here the overshoot parameter inferred earlier from our simulation of the DB-to-DQ transformation, but it is not clear that this choice is also suitable for the DA-to-DC transformation. In fact, the latter case probably requires a higher overshoot parameter, since the mixing episode occurs closer to the surface, where convective overshoot is expected to be more efficient. Finally, it would also be interesting to study the effect of hydrogen accretion on the chemical evolution.

\section{Summary and Conclusion} \label{sec:conclu}

The aim of this work was to investigate the theory of the spectral evolution of white dwarfs by carrying out numerical calculations of element transport in cooling white dwarfs. To do so, we relied on the STELUM software, which is the latest version of the stellar evolutionary code developed at the Universit\'e de Montr\'eal. First, we presented for the first time a thorough description of STELUM. We successively discussed the physical ingredients and numerical techniques used to model the structure and evolution of stars and the transport of chemical species. We emphasized that several transport mechanisms, among which all relevant types of atomic diffusion and convective mixing, are included as time-dependent diffusive processes and are fully coupled to the evolution. We hope that this paper can serve as a valuable reference for future studies using STELUM. Further details about the code can be found in Appendix \ref{sec:app_a} or provided upon request to the lead author.

Then, we presented the results of two sophisticated simulations of the chemical evolution of white dwarfs. In the first case, we modeled the PG 1159$-$DO$-$DB$-$DQ evolutionary channel from beginning to end. In the early evolution, the hot PG 1159 star initially retains a carbon- and oxygen-rich atmosphere due to a stellar wind, but then turns into a pure-helium DO white dwarf once mass loss ceases and gravitational settling becomes efficient. The resulting envelope consists of a double-layered structure: a pure-helium layer atop a PG 1159-like shell. Although the pure-helium layer grows with time, the double-layered structure persists throughout the cooler DB phase because of the long diffusion timescales at the base of the envelope. The superficial helium convection zone eventually reaches into the PG 1159-like shell, thereby bringing a small amount of carbon to the surface and transforming the DB star into a DQ star. At low effective temperatures, the dredged-up carbon partially sinks back below the convective region, and thus the atmospheric carbon abundance decreases with cooling. We demonstrated that our calculation is in excellent quantitative agreement with the well-established empirical relation between the carbon abundance and the effective temperature of cool DQ white dwarfs. We nevertheless pointed out that this portion of the evolution depends sensitively on several model parameters, the most uncertain of which is the extent of convective overshoot.

In the second case, we performed a full-fledged simulation of the DO$-$DA$-$DC evolutionary channel. At the outset, the helium-rich envelope of our DO white dwarf model contains a small amount of hydrogen. Gravitational settling causes this hydrogen to diffuse upward, a process that is initially very slow due to the competing effect of the stellar wind. A hydrogen layer gradually forms at the surface, eventually turning the DO star into a DA star. Later on, a superficial convection zone appears and thoroughly dilutes the hydrogen layer into the underlying helium envelope. Consequently, in terms of chemical composition, the final outcome is strikingly similar to the initial state: a helium-dominated atmosphere with an undetectable quantity of hydrogen, which corresponds to the DC spectral type. To our knowledge, this is the first time that this mixing event is computed in a time-dependent and self-consistent way. More quantitatively, for an initial hydrogen mass fraction $X_{\rm H} = 0.0001$, our simulated DO-to-DA and DA-to-DC transformations occur at $\Teff \sim 50,000$ and 10,000 K, respectively. Although these transition temperatures admittedly depend on the mass-loss and mixing prescriptions, they are both remarkably consistent with the observed variation of the numbers of hydrogen-rich and helium-rich white dwarfs along the cooling sequence. Moreover, we showed that the lower branch of the bifurcation seen in the Gaia color-magnitude diagram is perfectly matched by the cooling track of our final hydrogen-bearing DC star and can therefore be interpreted as the signature of convective mixing.

Clearly, the results presented in this paper are merely a glimpse of what can be learned from numerical simulations of element transport in cooling white dwarfs. Such calculations hold a huge potential for new advances in our theoretical understanding of spectral evolution, and this potential has yet to be fully exploited. An obvious step is the exploration of the model parameter space relevant to the two simulations discussed in Section \ref{sec:simu}, which could provide answers to a number of important questions. How does the carbon content of a PG 1159 star impact the carbon abundance of its DQ progeny? What is the effect of varying the hydrogen content of a hot DO white dwarf on the DO-to-DA and DA-to-DC transitions? What is the ultimate fate of a PG 1159-type object containing hydrogen (a so-called hybrid PG 1159 star)? How do fundamental parameters such as the total stellar mass and the envelope mass influence the transport of chemical species? What are the consequences of changing transport prescriptions such as the strength of the stellar wind, the efficiency of convective mixing, and the extent of convective overshoot? Finally, we note that several other unsettled problems related to the spectral evolution of white dwarfs need to be addressed in future research work. Notable examples include the origin of hydrogen in DBA stars, the carbon abundance pattern of massive DQ stars, and the interplay between the accretion, diffusion, and mixing of metals in DAZ, DBZ, and DZ white dwarfs.

\acknowledgments

We would like to acknowledge the essential contribution of our late colleague Gilles Fontaine to the development of the Montreal white dwarf evolutionary code. This work was supported by the Natural Sciences and Engineering Research Council (NSERC) of Canada and the Fonds de Recherche du Qu\'ebec $-$ Nature et Technologie (FRQNT). S.B. acknowledges support from the Laboratory Directed Research and Development program of Los Alamos National Laboratory (20190624PRD2) and from NSERC's Banting Postdoctoral Fellowship program.

\appendix

\section{Additional Details on the STELUM Code} \label{sec:app_a}

\subsection{Equations of Stellar Structure and Evolution} \label{sec:app_a_evol}

STELUM relies on the standard one-dimensional theory of stellar structure and evolution (see for instance \citealt{kippenhahn2012}). Consider a spherical, non-rotating, non-magnetic star of given total mass $M$ and chemical composition, the latter being specified by $I$ elemental mass fractions $X_1,X_2,...,X_I$ at each depth. The structure and evolution of such an object is governed by the well-known set of four differential equations describing mass conservation, hydrostatic equilibrium, energy conservation, and energy transport. In order to model the atmosphere as well, a fifth equation defining the optical depth must also be included. In Lagrangian form, these equations can be written as
\begin{align}
\frac{dr}{dm} &= \frac{1}{4 \pi r^2 \rho} , \label{eq:evol_1} \\[5pt]
\frac{dP}{dm} &= -\frac{Gm}{4 \pi r^4} , \\[5pt]
\frac{dl}{dm} &= \epsilon_{\rm nuc} - \epsilon_{\rm neu} + \epsilon_{\rm grav} , \\[5pt]
\frac{dT}{dm} &= -\frac{G m T}{4 \pi r^4 P} \nabla , \\[5pt]
\frac{d\tau}{dm} &= -\frac{\kappa}{4 \pi r^2} . \label{eq:evol_2}
\end{align}
In these expressions, the independent variables are the interior mass $m$ and time $t$ (which does not appear explicitly), whereas the five unknown variables are the radius $r$, total pressure $P$, luminosity $l$, temperature $T$, and Rosseland optical depth $\tau$. As for the remaining quantities, $G$ is the gravitational constant, $\rho$ is the mass density, $\epsilon_{\rm nuc}$, $\epsilon_{\rm neu}$, and $\epsilon_{\rm grav}$ are the rates of energy gain/loss per unit mass due to nuclear reactions, neutrino emission, and gravothermal processes, respectively, $\nabla = d \ln T / d \ln P$ is the logarithmic temperature gradient, and $\kappa$ is the Rosseland mean opacity. These functions must be specified a priori through an appropriate description of the microphysics of the stellar material, which is the subject of Section \ref{sec:code_evol}. Equations \ref{eq:evol_1}$-$\ref{eq:evol_2} are also supplemented with standard boundary conditions.

\subsection{Treatment of Convection} \label{sec:app_a_conv}

In a stellar model, convectively stable regions are characterized by the radiative temperature gradient, $\nabla_{\rm rad}$, given by Equation \ref{eq:rad_grad}. This is a slightly modified version of the standard expression that accounts for the breakdown of the diffusion approximation near the surface through the correction factor $W$. Convectively unstable layers are characterized instead by the convective temperature gradient, $\nabla_{\rm conv}$, which is evaluated using the mixing-length theory. In this case too, we adapt the usual equations to incorporate non-diffusive effects. We briefly describe our treatment of convection in the following.

Within the mixing-length formalism, it can be shown that the convective temperature gradient obeys a cubic polynomial equation that only depends on the local properties of the stellar material (see for instance \citealt{kippenhahn2012}). In order to write this expression compactly, we first define two dimensionless quantities,
\begin{align}
U &= \frac{c \sigma T^3}{W \rho^2 \kappa c_P \ell_{\rm conv}^2} \left( \frac{H_P}{a g} \frac{\chi_\rho}{\chi_T} \right)^{1/2} , \\[5pt]
V &= \frac{16 W}{3 b c} .
\end{align}
Recall that $\sigma$ is the Stefan-Boltzmann constant, $T$ is the temperature, $\rho$ is the density, $\kappa$ is the opacity, $c_P$ is the specific heat capacity at constant pressure, $\chi_T$ and $\chi_\rho$ are standard thermodynamic derivatives, $g$ is the gravitational acceleration, $H_P$ is the pressure scale height, $\ell_{\rm conv}$ is the mixing length, and $a$, $b$, and $c$ are numerical constants describing the geometry of the convective cells. It is in these two definitions that we include the effects of non-diffusive energy transfer through the correction factor $W$.

For purposes of numerical accuracy, the cubic polynomial equation for $\nabla_{\rm conv}$ is formulated differently in the limiting cases $U \gg 1$ and $U \ll 1$, which correspond to the regimes of low and high convective efficiency, respectively. For $U \gg 1$, we solve
\begin{equation}
x^3 + U V x^2 + U^2 V x - U V \left( \nabla_{\rm rad} - \nabla_{\rm ad} \right) = 0 ,
\end{equation}
where $x$ is related to $\nabla_{\rm conv}$ by
\begin{equation}
\nabla_{\rm conv} = \nabla_{\rm rad} - \frac{x^3}{U V} .
\end{equation}
For $U \ll 1$, we instead solve
\begin{equation}
x^3 + U \left( 2 V - 3 \right) x^2 + 3 U^2 x - 8 U V \left( \nabla_{\rm rad} - \nabla_{\rm ad} \right) - U^3 \left( 2 V + 1 \right) = 0 ,
\end{equation}
where $x$ is now related to $\nabla_{\rm conv}$ by
\begin{equation}
\nabla_{\rm conv} = \nabla_{\rm ad} + \frac{x^2 - U^2}{4} .
\end{equation}
As before, $\nabla_{\rm ad}$ denotes the adiabatic temperature gradient. Note that both forms are entirely equivalent for intermediate convective efficiencies. 

Once the convective temperature gradient is known, the convective velocity, which appears in Equation \ref{eq:coeff_conv} for the convective diffusion coefficient, can be computed as
\begin{equation}
v_{\rm conv} = \ell_{\rm conv} \left( \frac{a g}{H_P} \frac{\chi_T}{\chi_\rho} \right)^{1/2} \big[ U V \left( \nabla_{\rm rad} - \nabla_{\rm conv} \right) \big]^{1/3} .
\end{equation}

\subsection{Evaluation of the Hopf Function} \label{sec:app_a_hopf}

The Hopf function $H$ enters Equation \ref{eq:atmo_fac} for the atmospheric correction factor $W$ in the grey-atmosphere approximation. It is defined by the integral equation 
\begin{equation}
\tau + H(\tau) = \frac{1}{2} \int_0^\infty [ t + H(t) ] \ E_1 ( |t-\tau| ) \ dt ,
\end{equation}
where $\tau$ is the optical depth and $E_1$ is the so-called first exponential integral \citep{mihalas1978}. Thus, it cannot be expressed in a closed analytic form. To evaluate the Hopf function efficiently, we rely on a seventh-order polynomial fit to a numerical solution of the above equation. Our fit is given by 
\begin{equation}
H = \sum_{k=0}^{7} c_k x^k ,
\end{equation}
where $x$ is defined as
\begin{equation}
x = \frac{2}{\pi} \arcsin{ \big[ \exp{ \left( -\tau \right) } \big] } ,
\end{equation}
and the values of the coefficients are $c_0 = 0.7104460$, $c_1 = -0.02015790$, $c_2 = -0.08132497$, $c_3 = -0.3250189$, $c_4 = 0.8943672$, $c_5 = -1.1284420$, $c_6 = 0.5274319$, and $c_7 = 0.00004964$.

\subsection{Treatment of Nuclear Burning and Neutrino Emission} \label{sec:app_a_nuc}

Our treatment of nuclear burning and neutrino emission involves standard formulas (see for instance \citealt{kippenhahn2012}), which we explicitly provide here for completeness.

Nuclear burning appears in the equations of stellar structure in the form of a rate of energy generation per unit mass, which is given by
\begin{equation}
\epsilon_{\rm nuc} = \sum_{\mathcal N} \frac{R_{\mathcal N} Q_{\mathcal N}}{\rho} ,
\end{equation}
where, for some nuclear reaction $\mathcal N$, $R_{\mathcal N}$ is the reaction rate per unit volume and $Q_{\mathcal N}$ is the energy released by a single reaction (corrected for neutrino losses). The sum runs over all relevant reactions, and $\rho$ denotes the density as usual. Suppose that atomic species $i$ and $j$ are the reactants of reaction $\mathcal N$, which can thus be written symbolically as $\nu_i {\mathbb A}_i + \nu_j {\mathbb A}_j \rightarrow$ (products), where $\nu_i$ and $\nu_j$ are the stoichiometric coefficients and ${\mathbb A}_i$ and ${\mathbb A}_j$ are the chemical symbols. The reaction rate can then be expressed as
\begin{equation}
R_{\mathcal N} = \left( \frac{X_i}{A_i} \right)^{\nu_i} \left( \frac{X_j}{A_j} \right)^{\nu_j} \frac{\left( N_{\rm A} \rho \right)^{\nu_i + \nu_j}}{\nu_i! \nu_j!} f_{\mathcal N} \langle \sigma v \rangle_{\mathcal N} ,
\label{eq:rate_nuc}
\end{equation}
where $N_{\rm A}$ is Avogadro's number, $A_i$ and $A_j$ are the atomic weights, $X_i$ and $X_j$ are the mass fractions, $\langle \sigma v \rangle_{\mathcal N}$ is the reaction cross section, and $f_{\mathcal N}$ is the electron screening factor.

Energy losses due to the emission of neutrinos through leptonic reactions are considered in a similar fashion. The rate of energy loss per unit mass is given by
\begin{equation}
\epsilon_{\rm neu} = \sum_{\mathcal L} \frac{Q_{\mathcal L}}{\rho} ,
\end{equation}
where, for some leptonic neutrino-producing process $\mathcal L$, $Q_{\mathcal L}$ is the energy carried away per unit time and unit volume. Once again, the sum runs over all relevant processes. The physical quantities $Q_{\mathcal N}$, $\langle \sigma v \rangle_{\mathcal N}$, $f_{\mathcal N}$, and $Q_{\mathcal L}$ are taken from external sources listed in Section \ref{sec:code_evol}.

Nuclear burning also enters the equations governing element transport in the form of a source/sink term, $S_{{\rm nuc},i}$. This term is simply the net rate of change of the elemental mass fraction $X_i$ due to the combined effect of all reactions that add or remove particles of type $i$. If ${\mathcal N}_i^{\scriptscriptstyle +}$ and ${\mathcal N}_i^{\scriptscriptstyle -}$ denote reactions that create and destroy nuclei of type $i$ with stoichiometric coefficients $\nu_{{\mathcal N}_i^+}$ and $\nu_{{\mathcal N}_i^-}$, respectively, we can write
\begin{equation}
S_{{\rm nuc},i} = \frac{A_i}{N_{\rm A} \rho} \left( \sum_{{\mathcal N}_i^+} \nu_{{\mathcal N}_i^+} R_{{\mathcal N}_i^+} - \sum_{{\mathcal N}_i^-} \nu_{{\mathcal N}_i^-} R_{{\mathcal N}_i^-} \right) ,
\end{equation}
with the reactions rates given by Equation \ref{eq:rate_nuc}.

\vspace{2mm}
\section{Animations of Our Chemical Evolution Simulations} \label{sec:app_b}

\begin{figure*}[!h]
\centering
\vspace{-1mm}
\includegraphics[width=0.375\columnwidth,clip=true,trim=1.5cm 0.0cm 0.75cm 0.0cm,angle=270]{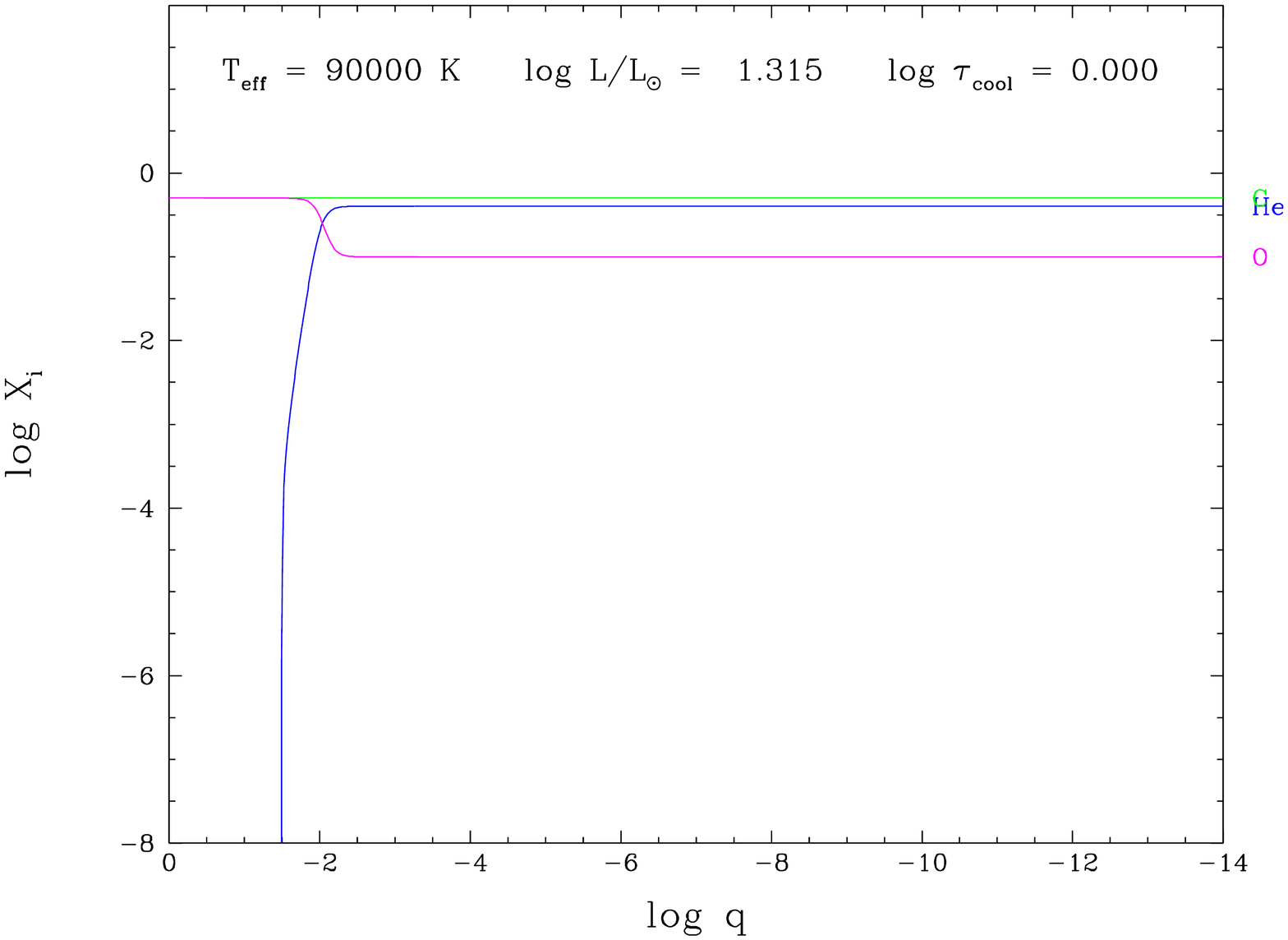}
\caption{Time evolution of the chemical structure, represented here as the run of the elemental mass fractions with fractional mass depth ($q = 1 - m/M$), throughout our entire PG 1159$-$DO$-$DB$-$DQ simulation (shown partially in Figure \ref{fig:evol_DO-DB-DQ}). The hydrogen, helium, carbon, and oxygen abundance profiles are displayed as red, blue, green, and magenta curves, respectively. The location of the base of the convection zone is indicated by a black dashed line. The effective temperature, total luminosity, and cooling age are given at the top of the graph. The animation is available in the online version of the article. The real-time duration of the video is 41 s.}
\label{fig:evol_DO-DB-DQ_0}
\end{figure*}

\begin{figure*}[!h]
\centering
\vspace{-1mm}
\includegraphics[width=0.375\columnwidth,clip=true,trim=1.5cm 0.0cm 0.75cm 0.0cm,angle=270]{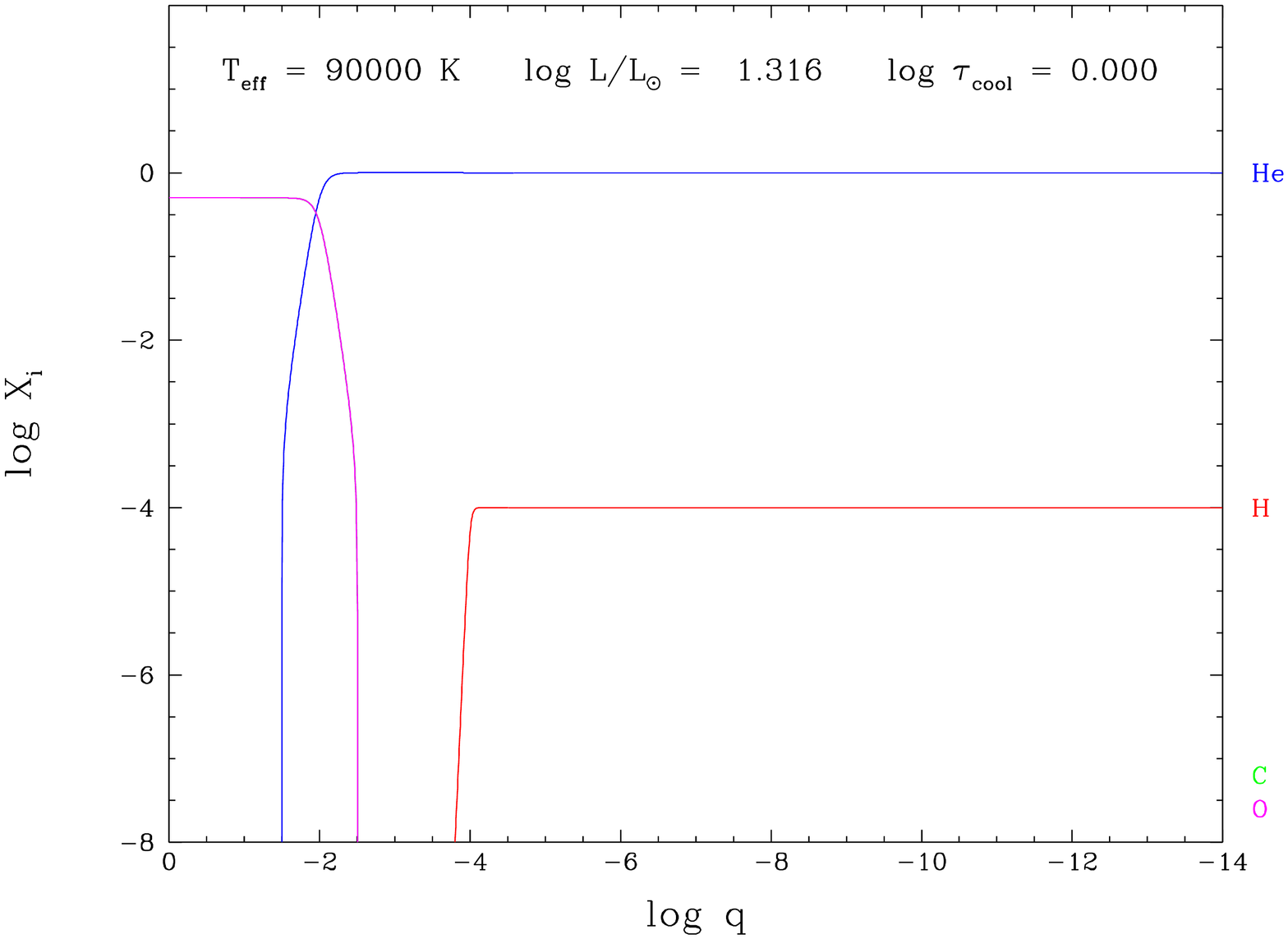}
\caption{Same as Figure \ref{fig:evol_DO-DB-DQ_0}, but for our DO$-$DA$-$DC simulation (shown partially in Figure \ref{fig:evol_DO-DA-DC}). The animation is available in the online version of the article. The real-time duration of the video is 43 s.}
\label{fig:evol_DO-DA-DC_0}
\end{figure*}

Figures \ref{fig:evol_DO-DB-DQ_0} and \ref{fig:evol_DO-DA-DC_0} host animated versions of Figures \ref{fig:evol_DO-DB-DQ} and \ref{fig:evol_DO-DA-DC}, respectively, in the online version of the article. These videos show the time evolution of the chemical profile throughout our whole PG 1159$-$DO$-$DB$-$DQ and DO$-$DA$-$DC simulations, discussed at length in Sections \ref{sec:simu_DO-DB-DQ} and \ref{sec:simu_DO-DA-DC}.

\bibliographystyle{aasjournal}
\bibliography{main}

\end{document}